\documentclass[twocolumn,a4paper]{article}
\usepackage[utf8]{inputenc}
\usepackage[T1]{fontenc}
\usepackage{amsmath,amssymb,amsfonts}
\usepackage{amsthm}
\usepackage{graphicx}
\usepackage{xcolor}
\usepackage{hyperref}
\usepackage{cite}
\usepackage{orcidlink}
\usepackage{geometry}
\usepackage{abstract}
\theoremstyle{plain}
\newtheorem{proposition}{Proposition}
\newtheorem{lemma}{Lemma}
\theoremstyle{remark}
\newtheorem*{remark}{Remark}

\title{\vspace{-1.5em}\textbf{Geodesic Dynamics, Static Spheres, and Aschenbach-Like Effect in Hairy Schwarzschild Black Holes}}
\author{\textbf{Mohammad Ali S. Afshar \orcidlink{0009-0001-3133-5992}}\textsuperscript{1,2,3}\thanks{m.a.s.afshar@gmail.com}, \textbf{J. Sadeghi \orcidlink{0000-0002-6764-7920}}\textsuperscript{1}\thanks{pouriya@ipm.ir}, and \textbf{H. Jafari}\textsuperscript{4}\thanks{jafarh@unisa.ac.za} \\[0.5em]
\small\textsuperscript{1}Department of Theoretical Physics, Faculty of Basic Sciences, University of Mazandaran, P. O. Box 47416-95447, Babolsar, Iran \\
\small\textsuperscript{2}School of Physics, Damghan University, P. O. Box 3671641167, Damghan, Iran \\
\small\textsuperscript{3}Center for Theoretical Physics, Khazar University, 41 Mehseti Street, Baku, AZ1096, Azerbaijan\\
\small\textsuperscript{4}Department of Mathematical Sciences, University of South Africa, UNISA 0003, Pretoria, South Africa}
\date{\today}
\newcommand{\keywords}[1]{\par\vspace{0.3em}\noindent\textbf{Keywords:} #1}

\begin{document}
\maketitle
\thispagestyle{empty}

\begin{abstract}
Static spheres, stable photon spheres and an Aschenbach-like increase of orbital angular velocity, reported for electromagnetic and massive-gravity black holes, are shown to occur in the neutral, asymptotically flat hairy Schwarzschild black hole of Extended Gravitational Decoupling (EGD): vacuum Schwarzschild deformed only by an exponentially localised geometric term. For $g_{tt}g_{rr}=-1$, an exact identity (specialising known results) relates $g_{tt}'$ to the integral of the effective tangential pressure, tying static spheres to the Strong Energy Condition (SEC), which the effective source violates for $r<2M-\ell_0$. For $M=\ell_0=1$, $\alpha=7.8$: three photon spheres and a static-sphere pair outside the horizon, with computed topological charges; three disconnected timelike-circular-orbit branches with a single marginally stable orbit, at $r=5.915$. For $\ell_0\le M$, closed-form conditions give the static-sphere pair at $\alpha=e^{2}M/(2M-\ell_0)$ and the additional photon spheres within a window bounded by the roots of a cubic, defining four regimes. The angular velocity increases with radius wherever $r\mathcal A''>\mathcal A'$, requiring positive effective tangential pressure: next to every stable photon sphere and stable static sphere and, for sufficiently large $\alpha$, also where neither exists; static spheres without a stable photon sphere give an interior maximum on a single connected branch. Test-particle limit throughout.
\end{abstract}


\section{Introduction}
Black hole spacetimes often display orbital and geometric features that parallel phenomena across different theoretical frameworks~\cite{1,2,3,4,5}. In the Kerr geometry the Aschenbach effect is a non-monotonic radial profile of the orbital velocity of circular geodesics measured by zero-angular-momentum observers (ZAMOs), which appears for spins $a/M>0.9953$~\cite{6}. For a distant observer, the angular velocity of timelike circular orbits decreases monotonically with radius in the Schwarzschild and Kerr geometries. An analogous behaviour was nevertheless found in static, non-rotating black holes: the angular velocity $\Omega$ measured with respect to a distant static observer increases with radius over part of the family of circular orbits. A stable photon sphere outside the horizon, together with static spheres, has been identified as the indicator of this behaviour~\cite{7,8,9}. It has been reported for dyonic black holes with nonlinear electrodynamics and for black holes in massive gravity~\cite{7,8,9,10,11}.

A related geodesic phenomenon in static spacetimes is the existence of static spheres---hypersurfaces on which massive test particles remain at rest ($\Omega=0$) relative to asymptotic static observers~\cite{13}. Static points of this kind were first found in the equatorial plane of rotating non-Kerr black holes~\cite{12}; in a static, spherically symmetric spacetime they extend to complete spheres~\cite{13}. For a metric with $g_{tt}g_{rr}=-1$, the Einstein equations give an exact relation between $g_{tt}'$ and the effective tangential pressure (Eq.~\eqref{eq:komar} below). It is the form, for $g_{tt}g_{rr}=-1$, of the result of Cvetič, Gibbons, and Pope that the SEC requires $|g_{tt}|$ to increase monotonically outside a regular horizon~\cite{3_2,2_2}; static spheres therefore require a violation of the SEC of the effective source~\cite{13,song}. These authors also found that the SEC-respecting asymptotically flat solutions they examined possess one and only one photon sphere, and conjectured that a violation of the SEC or of the DEC is necessary for a stable photon sphere outside the horizon~\cite{3_2}. Counterexamples with $g_{tt}g_{rr}\neq-1$ have been found~\cite{2_2}. Static spheres have been documented in non-linear electrodynamics models~\cite{5_2,13} and in neutral black holes within massive gravity~\cite{11}. A local analysis of the energy conditions at static spheres has recently been given in Ref.~\cite{song}.

Static spheres have also been discussed as theoretical platforms for stationary energy-harvesting structures or Dyson-like configurations around compact objects~\cite{13}. However, as emphasized by Hod~\cite{12_2,13_2}, a physical shell or megastructure possesses self-gravity, which alters the background metric and shifts the equilibrium radius away from the test-particle static sphere. Furthermore, spacetimes admitting stable photon spheres may be subject to long-lived perturbation modes and potential non-linear instabilities~\cite{14_2}. A complete evaluation of structural or spacetime stability therefore requires specifying a fundamental action for the gravitational source.

In this work we examine the geodesic structure of a hairy Schwarzschild black hole (H-SBH) constructed via the Extended Gravitational Decoupling (EGD) approach~\cite{57,59,60,61}. This solution is neutral, has the vacuum Schwarzschild solution as its seed, keeps the ADM mass $M$, and differs from it only by an exponentially localised term of range $\mathcal M=M-\ell_0/2$. Static spheres, a stable photon sphere, and an Aschenbach-like increase of $\Omega$ have been reported for quasi-topological electromagnetism~\cite{5_2,13,7} and for dRGT massive gravity~\cite{11}; in the massive-gravity examples of Ref.~\cite{11} the lapse function contains a $\Lambda r^{2}/3$ term with $\Lambda\neq0$. General statements on the pairing of static spheres and on the local sign of the SEC combination at each member of a pair are available~\cite{3_2,13,song}. Photon spheres, shadows, accretion images, wave echoes, and absorption spectra of the H-SBH have been studied in Refs.~\cite{2_2,4_2,6_2,7_2,8_2}; in particular, Ref.~\cite{2_2} mapped the region of the $(\alpha,\ell_0/M)$ plane with double photon spheres and found that in it the DEC and SEC are violated while the NEC and WEC can be respected. We ask whether a purely geometric, neutral, and asymptotically flat realisation of static spheres and of the Aschenbach-like behaviour exists, and what it looks like. In Sec.~\ref{sec:hsbh} we determine, for $\ell_0=M$ and $\alpha=7.8$, the photon spheres, static spheres, topological charges, branches of circular orbits, and the SEC bound; the effective density and pressures are recomputed directly from the Einstein tensor (Appendices~A and~B). In Sec.~\ref{sec:param} we derive closed-form conditions in the $(\ell_0/M,\alpha)$ plane for the creation of the static-sphere pair and of the additional photon spheres, classify the resulting regimes, relate the criterion for an increasing $\Omega$ to the effective tangential pressure, and show that the increase also occurs in a part of the parameter space with neither a static sphere nor a stable photon sphere. The relation to the SEC is of two kinds. At an unstable static sphere the SEC violation is local, since $p_{t,\mathrm{eff}}=\mathcal A''/16\pi<0$ there. At a stable static sphere $p_{t,\mathrm{eff}}>0$, and the equilibrium is the outcome of the integrated effect of Eq.~\eqref{eq:komar}. For the photon sphere at $r=0.7110$ the SEC violation is a feature of this particular solution rather than a general necessity, since black holes with multiple photon spheres satisfying all four energy conditions are known to exist~\cite{2_2}; the known examples have $g_{tt}g_{rr}\neq-1$, whereas the examples with $g_{tt}g_{rr}=-1$ studied in Ref.~\cite{2_2} violate at least one energy condition.

\section{Key Concepts and Definitions}

\subsection{Static Spheres and the Strong Energy Condition}
\label{sec:static_sec}
For the class $g_{tt}g_{rr}=-1$ the energy-condition constraints on static spheres take a simple explicit form. We use the following relations, which follow from results of Refs.~\cite{3_2,13,song}: the monotonicity of $|g_{tt}|$ under the SEC~\cite{3_2}, the pairing of static spheres~\cite{13,song}, and the local sign of the SEC combination at each member of a pair~\cite{song}.

Consider a static, spherically symmetric spacetime with $g_{tt}g_{rr}=-1$, $\mathrm{d}s^2=-\mathcal A\,\mathrm{d}t^2+\mathcal A^{-1}\mathrm{d}r^2+r^2\mathrm{d}\Omega^2$. A static sphere of radius $r_{\mathrm{sp}}$ is a surface on which a massive test particle remains at rest ($\Omega=0$) relative to asymptotic static observers~\cite{13}; this requires $\mathcal A'(r_{\mathrm{sp}})=0$. We interpret the Einstein tensor as an effective source, $G^{\mu}{}_{\nu}=8\pi T^{\mu}{}_{\nu}$, with $T^{t}{}_{t}=-\rho$, $T^{r}{}_{r}=p_r$, and $T^{\theta}{}_{\theta}=T^{\phi}{}_{\phi}=p_t$. For this metric form $G^{t}{}_{t}=G^{r}{}_{r}$, hence $\rho+p_r=0$ identically, while $G^{\theta}{}_{\theta}=(r\mathcal A''+2\mathcal A')/(2r)$. The latter equation can be written as
\begin{equation}
\frac{\mathrm{d}}{\mathrm{d}r}\big(r^{2}\mathcal A'\big)=16\pi r^{2}p_{t},
\label{eq:komar}
\end{equation}
and integrating from a non-extremal horizon gives
\begin{equation}
r^{2}\mathcal A'(r)=r_h^{2}\mathcal A'(r_h)+16\pi\!\int_{r_h}^{r}\! r'^{2}p_{t}(r')\,\mathrm{d}r' .
\label{eq:komar_int}
\end{equation}
Since $\rho+p_r=0$, the SEC for this source consists of $p_t\ge0$ (from $\rho+p_r+2p_t\ge0$) and $\rho+p_t\ge0$. If $p_t\ge0$ for all $r>r_h$, Eq.~\eqref{eq:komar_int} gives $\mathcal A'>0$ outside the horizon: $|g_{tt}|$ increases monotonically and no static sphere exists. This is the property of SEC-respecting black holes established in Ref.~\cite{3_2} and used in Ref.~\cite{2_2}; Eq.~\eqref{eq:komar_int} is its explicit form for $g_{tt}g_{rr}=-1$. Three consequences follow.
\begin{enumerate}
\item A static sphere at $r_{\mathrm{sp}}$ requires $p_t<0$ on part of $(r_h,r_{\mathrm{sp}})$, such that the integral in Eq.~\eqref{eq:komar_int} equals $-r_h^{2}\mathcal A'(r_h)/16\pi$.
\item Since $\mathcal A$ grows from zero at the horizon, the innermost static sphere is a local maximum of $\mathcal A$ ($\mathcal A''\le0$), unstable if the zero of $\mathcal A'$ is simple.
\item At any static sphere $\mathcal A'=0$, so Eq.~\eqref{eq:komar} gives
\begin{equation}
p_t(r_{\mathrm{sp}})=\frac{\mathcal A''(r_{\mathrm{sp}})}{16\pi}.
\label{eq:pt_at_sp}
\end{equation}
It is negative at an unstable static sphere ($\mathcal A''<0$) and positive at a stable one ($\mathcal A''>0$).
\end{enumerate}
The SEC is therefore violated locally at an unstable static sphere. At a stable static sphere $p_t>0$, so the first SEC inequality holds locally; the second inequality, $\rho+p_t\ge0$, has to be checked separately for a given source (it holds for the source of Sec.~\ref{sec:hsbh}). The static equilibrium then results from the integrated effect expressed by Eq.~\eqref{eq:komar_int}.

\subsection{Aschenbach-Like Effect}
In the Kerr geometry the Aschenbach effect is a non-monotonic radial profile of the orbital velocity of timelike circular orbits (TCOs) measured by ZAMOs, which appears for rapidly spinning black holes ($a/M > 0.9953$)~\cite{6,25}. The angular velocity with respect to a distant observer decreases monotonically with radius in the Schwarzschild and Kerr geometries~\cite{7}. In static spherically symmetric geometries the analogous behaviour is defined through the angular velocity $\Omega=\mathrm{d}\phi/\mathrm{d}t$ of TCOs with respect to a distant static observer: the Aschenbach-like effect is present when $\Omega$ increases with $r$ over part of the TCO family~\cite{7,11}. Static spheres and a stable photon sphere outside the horizon have been identified as indicators of this behaviour~\cite{7}. The Kerr effect originates from frame dragging, whereas no frame dragging is present here: the effect is controlled by the radial structure of the metric function $\mathcal A(r)$.

\section{Methodology}

\subsection{Geometric Framework}
We consider a four-dimensional, static, spherically symmetric Lorentzian spacetime with line element:
\begin{equation}
\mathrm{d}s^{2} = -\mathcal{A}(r)\mathrm{d}t^{2} + \mathcal{A}^{-1}(r)\mathrm{d}r^{2} + r^{2}\big(\mathrm{d}\theta^{2} + \sin^{2}\theta \mathrm{d}\phi^{2}\big),
\end{equation}
where $\mathcal{A}(r)$ is the metric function, geometrized units ($\hbar=G=c=1$) are used, and the horizon sits at $\mathcal{A}(r_h)=0$. Writing $g_{rr}=\mathcal{A}^{-1}(r)$ rather than an independent function is a simplifying choice, not a general feature of static spacetimes; it holds for the EGD solution considered here and lets the same function $\mathcal{A}(r)$ fix both the horizon and the causal structure. The metric is also deliberately non-rotating, so that the orbital features studied below arise from the radial profile of $\mathcal{A}(r)$ alone, without any admixture of frame-dragging. Spherical symmetry permits restriction to the equatorial plane ($\theta=\pi/2$). Conserved energy $\mathcal{E}$ and angular momentum $\mathcal{L}_z$ per unit rest mass are given by:
\begin{equation}
\mathcal{E} = \mathcal{A}(r)\dot{t},\qquad \mathcal{L}_{z} = r^{2}\dot{\phi}.
\end{equation}

\subsection{Geodesics, Circular Orbits, and Static Spheres}
For timelike geodesics, the radial motion obeys $\dot{r}^{2} + \mathcal{V}_{\mathrm{eff}}^{(\mathrm{TL})} = 0$, with
\begin{equation}
\mathcal{V}_{\mathrm{eff}}^{(\mathrm{TL})} = \mathcal{A}(r)\left[1 + \frac{\mathcal{L}_{z}^{2}}{r^{2}}\right] - \mathcal{E}^{2}.
\end{equation}
Timelike circular orbits (TCO) require $\mathcal{V}_{\mathrm{eff}}^{(\mathrm{TL})} = \partial_{r} \mathcal{V}_{\mathrm{eff}}^{(\mathrm{TL})} = 0$. With
\begin{equation}
D(r)\equiv 2\mathcal{A}(r)-r\mathcal{A}'(r)
\end{equation}
this yields
\begin{equation}
\mathcal{E}_{\mathrm{CO}}^{2} = \frac{2\mathcal{A}^{2}}{D},\qquad \mathcal{L}_{z,\mathrm{CO}}^{2} = \frac{r^{3}\mathcal{A}'}{D},
\label{eq:E_L_CO}
\end{equation}
i.e.\ $\mathcal{E}_{\mathrm{CO}}=\sqrt{2}\,\mathcal{A}/\sqrt{D}$ and $\mathcal{L}_{z,\mathrm{CO}}=r^{2}\sqrt{\mathcal{A}'/r}/\sqrt{D}$. These reduce to the familiar Schwarzschild expressions when $\mathcal{A}(r)=1-2M/r$. Their ratio gives the orbital angular velocity with respect to the coordinate time, i.e.\ as measured by a distant static observer,
\begin{equation}
\Omega_{\mathrm{CO}}=\frac{\mathcal{A}\,\mathcal{L}_{z,\mathrm{CO}}}{r^{2}\mathcal{E}_{\mathrm{CO}}}=\sqrt{\frac{\mathcal{A}'(r)}{2r}} .
\label{eq:Omega_CO}
\end{equation}
Real values of $\mathcal{E}_{\mathrm{CO}}$ and $\mathcal{L}_{z,\mathrm{CO}}$ require both $D>0$ and $\mathcal{A}'\ge0$: if $D<0$ then $\mathcal{E}_{\mathrm{CO}}^{2}<0$, and if $\mathcal{A}'<0$ then either $\mathcal{L}_{z,\mathrm{CO}}^{2}<0$ (for $D>0$) or $\mathcal{E}_{\mathrm{CO}}^{2}<0$ (for $D<0$). The TCO therefore exist exactly where
\begin{equation}
D(r)>0\quad\text{and}\quad \mathcal{A}'(r)\ge0 .
\label{eq:tco_domain}
\end{equation}
The first condition is equivalent to $\beta\equiv-g_{tt}-g_{\phi\phi}\Omega^{2}=\mathcal{A}-r^{2}\Omega^{2}=D/2>0$.

Static spheres correspond to zero angular momentum ($\mathcal{L}_{z}=0$), which leads directly to:
\begin{equation}
\mathcal{A}^{\prime}(r_{\mathrm{sp}}) = 0,\qquad \mathcal{E}_{\mathrm{sp}} = \sqrt{\mathcal{A}(r_{\mathrm{sp}})}.
\label{eq:static_cond}
\end{equation}
For $\mathcal L_z=0$ the radial potential is $\mathcal A-\mathcal E^{2}$, so stability is governed by $\mathcal{A}''(r_{\mathrm{sp}}) > 0$ (stable) or $\mathcal{A}''(r_{\mathrm{sp}}) < 0$ (unstable). Static spheres are the end points of TCO branches, since $\Omega_{\mathrm{CO}}\to0$ there.

For null geodesics the same reduction gives $\dot{r}^{2} + \mathcal{L}_z^{2}\mathcal{V}_{\mathrm{eff}}^{(\mathrm{N})}(r) = \mathcal{E}^{2}$, with $\mathcal{V}_{\mathrm{eff}}^{(\mathrm{N})}(r) = \mathcal{A}(r)/r^{2}$. Circular photon orbits sit at extrema of $\mathcal{V}_{\mathrm{eff}}^{(\mathrm{N})}$, i.e.,
\begin{equation}
D(r)=2\mathcal{A}(r) - r\mathcal{A}'(r) = 0,
\label{eq:photon_sphere_cond}
\end{equation}
a maximum corresponding to an unstable photon sphere and a minimum to a stable one. This is the same combination that appears in Eq.~\eqref{eq:E_L_CO}: $\mathcal{E}_{\mathrm{CO}}$ and $\mathcal{L}_{z,\mathrm{CO}}$ diverge as a root of Eq.~\eqref{eq:photon_sphere_cond} is approached from the side where $D>0$, so a TCO branch ends at a photon sphere. While Eq.~\eqref{eq:photon_sphere_cond} already locates the photon spheres, it does not on its own indicate how many roots to expect or how to assign them a stability index in a way that generalizes to static spheres; for this we turn to a topological description, following Refs.~\cite{53,54} and~\cite{13}.

The stability of a TCO is governed by $V''_{\mathrm{eff}}\equiv\partial_r^2\mathcal{V}_{\mathrm{eff}}^{(\mathrm{TL})}$ evaluated at $(\mathcal E_{\mathrm{CO}},\mathcal L_{z,\mathrm{CO}})$,
\begin{equation}
V''_{\mathrm{eff}}=\frac{2}{D}\Big(\mathcal{A}\mathcal{A}''-2\mathcal{A}'^{2}+\frac{3\mathcal{A}\mathcal{A}'}{r}\Big),
\label{eq:Vpp}
\end{equation}
whose explicit form in terms of the functions $\varepsilon_i$ is given in Appendix~C. A TCO is stable if $V''_{\mathrm{eff}}>0$. The marginally stable circular orbit (MSCO) is a zero of $V''_{\mathrm{eff}}$ \emph{on the TCO domain}~\eqref{eq:tco_domain}; zeros of Eq.~\eqref{eq:Vpp} outside this domain are formal and carry no physical meaning. As $\mathcal L_z\to0$ at a static sphere, $V''_{\mathrm{eff}}\to\mathcal A''(r_{\mathrm{sp}})$, consistent with the static-sphere criterion above.

\subsection{Topological Charge of Photon Spheres and Static Spheres}
To characterize photon orbits, we employ the regularized vector field $\mathbf{n} = (n^r, n^\theta) = \phi / \|\phi\|$ defined on the $(r,\theta)$ plane, where $\phi^r = \sqrt{g_{rr}}\,\partial_r \mathcal{H}$, $\phi^\theta = \sqrt{g_{\theta\theta}}\,\partial_\theta \mathcal{H}$, and $\mathcal{H}(r,\theta) = \sqrt{\mathcal{A}(r)}/(r\sin\theta)$~\cite{53,54}. Multiplying the two components by arbitrary positive functions, for instance passing to the orthonormal-frame components, changes neither the zeros nor their winding numbers. Zeros of $\phi$ identify photon spheres. The topological charge $w_i$ of each zero is computed via the winding number, in the $(r,\theta)$ orientation:
\begin{equation}
w_i = \frac{1}{2\pi} \oint_{C_i} \epsilon_{ab} n^a \partial_\tau n^b \mathrm{d}\tau.
\end{equation}
Similarly, static spheres are classified using the vector field components $\phi^r = \mathcal{A}'(r)$ and $\phi^\theta = -\cos\theta/\sin^2\theta$~\cite{13}, i.e.\ $\phi=\nabla\Phi$ with $\Phi=\mathcal A(r)+1/\sin\theta$. At a zero on the equator the Jacobian of $\phi$ equals $\mathcal A''(r_{\mathrm{sp}})$, so the charge of a static sphere is $w=\mathrm{sign}\,\mathcal A''(r_{\mathrm{sp}})$: $-1$ for an unstable and $+1$ for a stable static sphere. The total charge of the static spheres of an asymptotically flat black hole vanishes, $\sum w_i=0$, so that static spheres come in pairs~\cite{13}.

\section{Hairy Schwarzschild Black Hole: Geodesic Analysis and SEC Bound}
\label{sec:hsbh}

The metric function of the hairy Schwarzschild black hole generated by EGD is given by~\cite{57}:
\begin{equation}
\mathcal{A}(r) = 1 - \frac{2M}{r} + \alpha \exp\left[-\frac{r}{M - \ell_0/2}\right],
\label{eq:metric_HSBh}
\end{equation}
where $M$ is the ADM mass, $\alpha$ is the deformation parameter, and $\ell_0$ is the primary hair length scale. Defining $\mathcal{M} \equiv M - \ell_0/2$, the metric reads $\mathcal{A}(r) = 1 - 2M/r + \alpha e^{-r/\mathcal{M}}$. Throughout we assume $\alpha>0$ and $\mathcal M>0$ (i.e.\ $\ell_0<2M$), so that the deformation decays exponentially and $M$ is indeed the ADM mass.

\subsection{Energy Conditions and SEC Violation Domain}
\label{sec:sec}
Write $h(r)\equiv\alpha e^{-r/\mathcal{M}}$. The Schwarzschild part of $\mathcal A$ is a vacuum solution, so the Einstein tensor, and hence the effective source $T^{\rm eff}_{\mu\nu}=\alpha\theta_{\mu\nu}$ of the EGD construction, is determined by $h$ alone (Appendix~A). With the signature $(-+++)$ used here, $\rho_{\mathrm{eff}}=-G^{t}{}_{t}/8\pi$, $p_{r,\mathrm{eff}}=G^{r}{}_{r}/8\pi$, and $p_{t,\mathrm{eff}}=G^{\theta}{}_{\theta}/8\pi$, which gives
\begin{align}
\rho_{\mathrm{eff}} &= \frac{\alpha e^{-r/\mathcal{M}}}{8\pi \mathcal{M} r^2} (r - \mathcal{M}), \\
p_{r,\mathrm{eff}} &= -\rho_{\mathrm{eff}}, \\
p_{t,\mathrm{eff}} &= \frac{\alpha e^{-r/\mathcal{M}}}{16\pi \mathcal{M}^2 r} \left(r - 2M + \ell_0\right). \label{eq:pt_correct}
\end{align}
(In the mostly-minus signature often used in the EGD literature, where the effective source is $\alpha\theta_{\mu\nu}$, the same quantities read $\rho_{\mathrm{eff}}=\alpha\theta^{0}{}_{0}$ and $p_{r,t}=-\alpha\theta^{1,2}{}_{1,2}$, up to the factor $8\pi$ in the normalisation of the coupling.) Since $\rho_{\mathrm{eff}}+p_{r,\mathrm{eff}}=0$ identically, the SEC reduces to $p_{t,\mathrm{eff}}\ge0$ together with $\rho_{\mathrm{eff}}+p_{t,\mathrm{eff}}\ge0$. From Eq.~\eqref{eq:pt_correct} the first condition is
\begin{equation}
r \ge 2M - \ell_0. \label{eq:sec_bound}
\end{equation}
The second condition, $\rho_{\mathrm{eff}}+p_{t,\mathrm{eff}}=\alpha e^{-r/\mathcal M}(r^{2}-2\mathcal M^{2})/(16\pi\mathcal M^{2}r^{2})\ge0$, holds for $r\ge\sqrt2\,\mathcal M$, which is implied by Eq.~\eqref{eq:sec_bound} because $2\mathcal M>\sqrt2\,\mathcal M$. Hence the SEC holds if and only if $r\ge2M-\ell_0$, and for $r<2M-\ell_0$ we have $p_{t,\mathrm{eff}}<0$: the SEC is violated. Outside the horizon this requires $r_h<2M-\ell_0$. For the parameter choice $M=1$, $\ell_0=1$, and $\alpha=7.8$ ($\mathcal{M}=0.5$) the horizon is at $r_h=0.5830$, the SEC is satisfied for $r \ge 1.0$ and violated for $r < 1.0$.

\begin{figure}[htbp]
    \centering
    \includegraphics[width=0.95\linewidth]{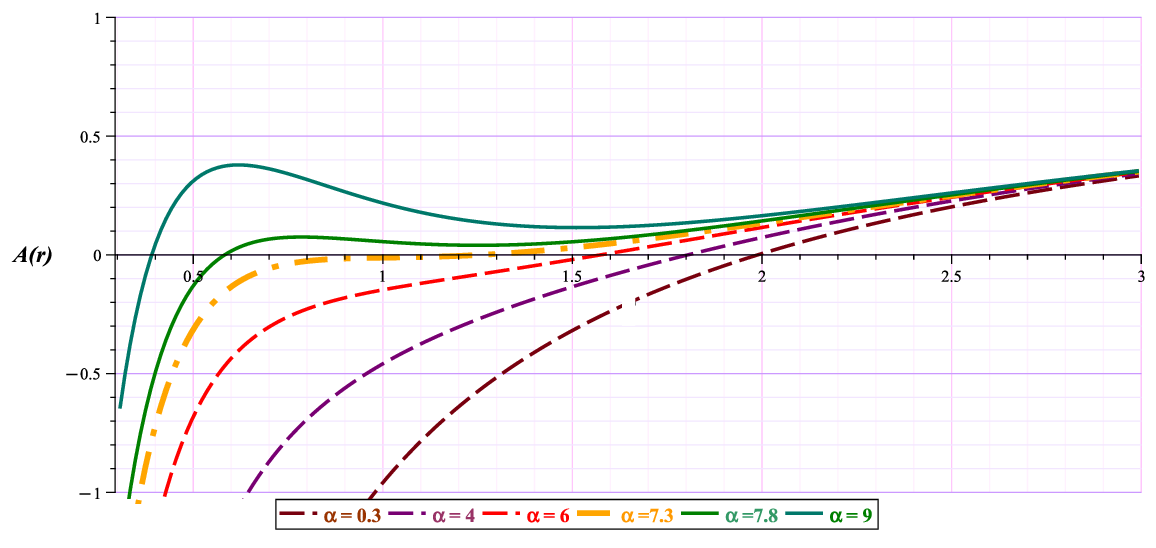}
    \caption{Metric function $\mathcal{A}(r)$ for $M=\ell_0=1$ and $\alpha=0.3,\,4,\,6,\,7.3,\,7.8,\,9$. The curves with $\alpha<e^{2}\simeq7.389$ have a single photon sphere outside the horizon; $\alpha=7.8$ and $\alpha=9$ lie in the window $e^{2}<\alpha\lesssim9.26$ with three photon spheres and a pair of static spheres.}
    \label{fig:metric_function}
\end{figure}

For $M=\ell_0=1$ the SEC holds everywhere outside the horizon for $\alpha<e^{2}\simeq7.389$, because then $r_h>1=2M-\ell_0$ (Fig.~\ref{fig:metric_function}). For $\alpha>e^{2}$ the horizon lies below $r=1$, the SEC is violated outside it, and $\mathcal A$ develops a local maximum and a local minimum, which give a pair of static spheres. The three-photon-sphere window is $e^{2}<\alpha<\alpha_{\rm hi}\simeq9.26$ (Sec.~\ref{sec:param}); we fix $\alpha=7.8$, which lies inside it, at about $5.6\%$ above its lower edge.

\subsection{Photon Sphere Topology}
\begin{figure}[htbp]
    \centering
    \includegraphics[width=0.82\linewidth]{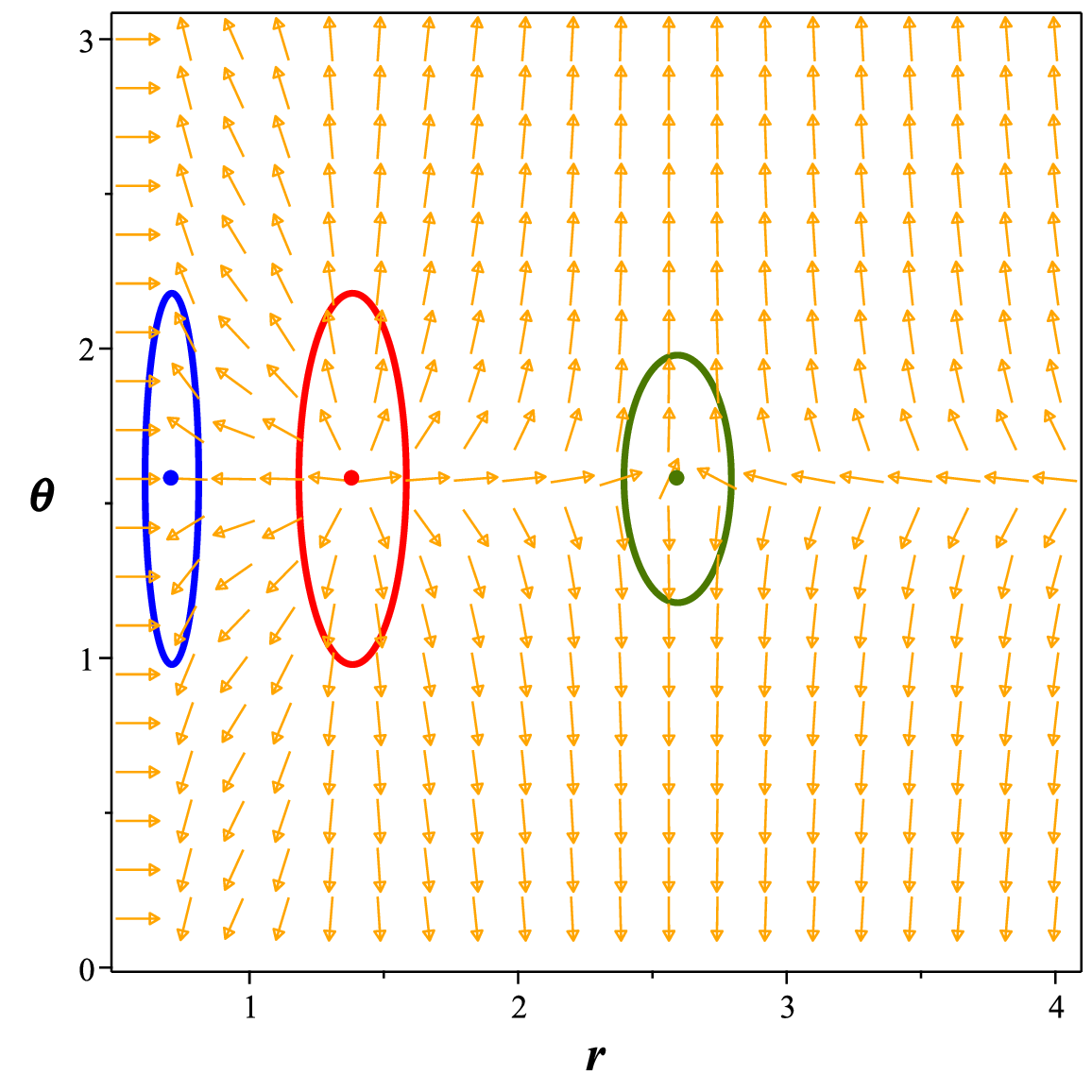}
    \hfill
    \includegraphics[width=0.82\linewidth]{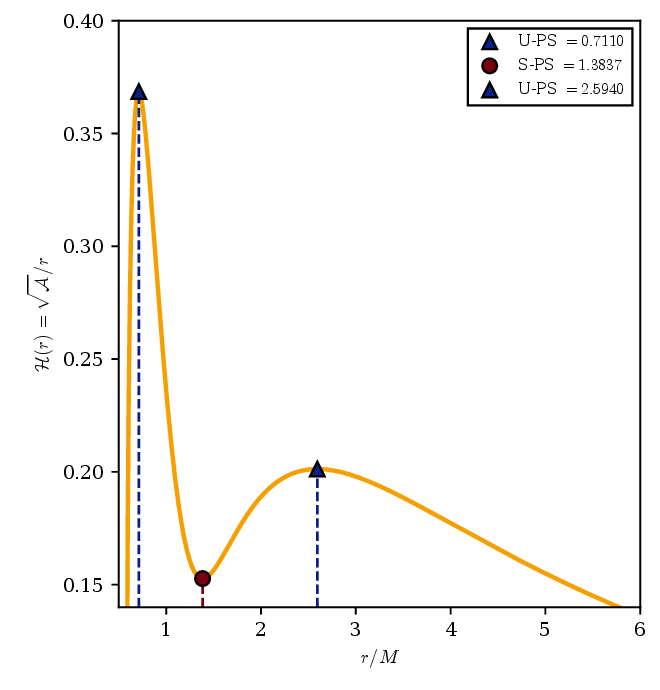}
    \caption{(a) The photon spheres at $(r,\theta)=(0.7110,\pi/2)$, $(1.3837,\pi/2)$, and $(2.5940,\pi/2)$ (unstable, stable, unstable) in the $(r,\theta)$ plane of the normal vector field $\mathbf{n}$, for $M=\ell_0=1$, $\alpha=7.8$. (b) The topological potential $\mathcal{H}(r)=\sqrt{\mathcal A}/r$ at $\theta=\pi/2$, with local maxima/minimum at the unstable/stable photon spheres.}
    \label{fig:photon_topology}
\end{figure}

Evaluating the vector field $\mathbf{n}$ for $M=1, \ell_0=1, \alpha=7.8$ yields three photon spheres (Fig.~\ref{fig:photon_topology}):
\begin{itemize}
\item Unstable photon sphere (U-PS1): $r = 0.7110$ ($w = -1$),
\item Stable photon sphere (S-PS): $r = 1.3837$ ($w = +1$),
\item Unstable photon sphere (U-PS2): $r = 2.5940$ ($w = -1$).
\end{itemize}
The total topological charge is $\sum w_i = -1$, consistent with black hole topology~\cite{53,54}. We note that the innermost photon sphere at $r = 0.7110$ lies inside the SEC-violating region ($r < 1.0$).

\subsection{Timelike Circular Orbits and Aschenbach-Like Effect}
\label{sec:tco}
\begin{figure*}[t]
    \centering
    \includegraphics[width=0.88\textwidth]{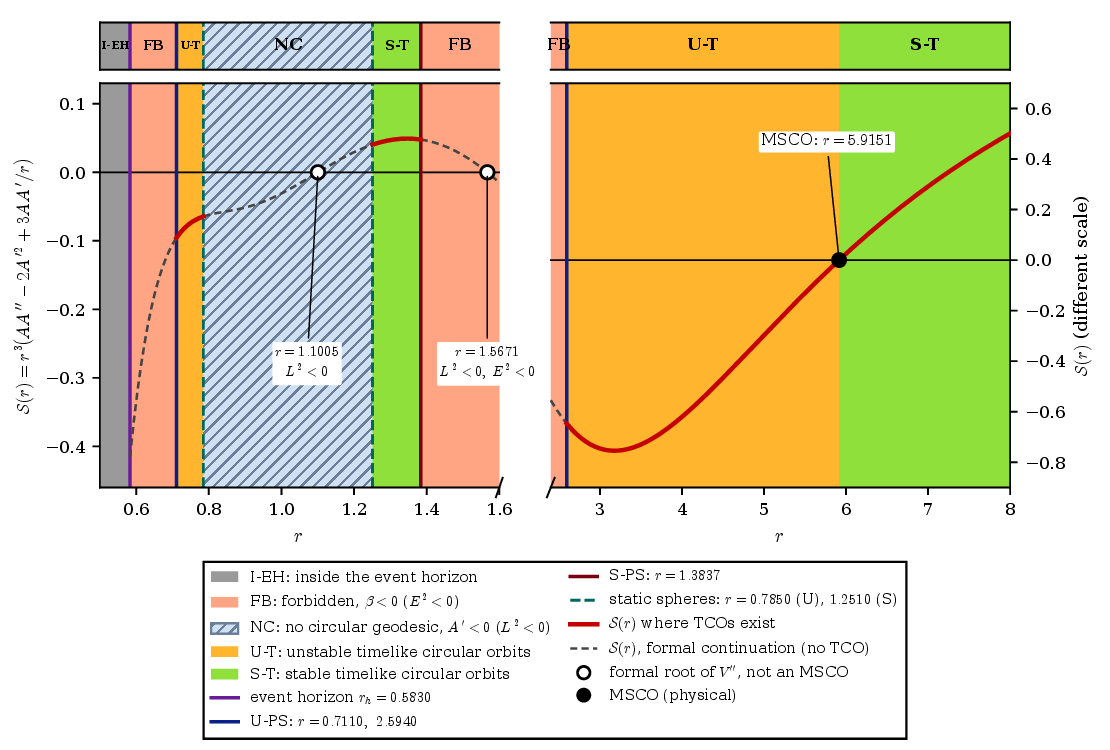}
    \caption{Domains of circular orbits for $M=\ell_0=1$, $\alpha=7.8$ (broken $r$ axis; the two panels have different vertical scales). Shaded bands: I-EH, inside the event horizon; FB, forbidden region $\beta<0$; NC, no circular geodesics ($\mathcal A'<0$); U-T and S-T, unstable and stable TCOs. The curve $\mathcal S(r)=r^{3}(\mathcal A\mathcal A''-2\mathcal A'^{2}+3\mathcal A\mathcal A'/r)$ has the sign of $V''_{\mathrm{eff}}=2\mathcal S/(r^{3}D)$ on the TCO domain, where $D>0$; it is drawn solid where TCOs exist and dashed elsewhere. The open circles are the zeros of $V''_{\mathrm{eff}}$ at $r=1.1005$ and $1.5671$, which lie where no TCO exists. The only MSCO is the filled circle at $r=5.9151$.}
    \label{fig:msco}
\end{figure*}

\begin{figure}[htbp]
    \centering
    \includegraphics[width=0.85\linewidth]{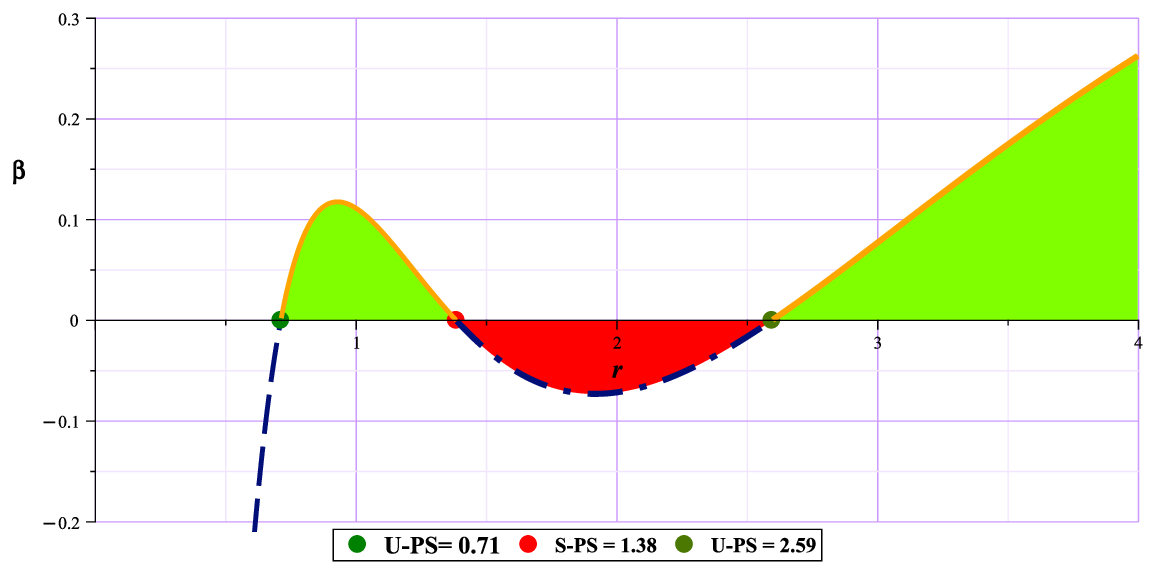}
    \hfill
    \includegraphics[width=0.85\linewidth]{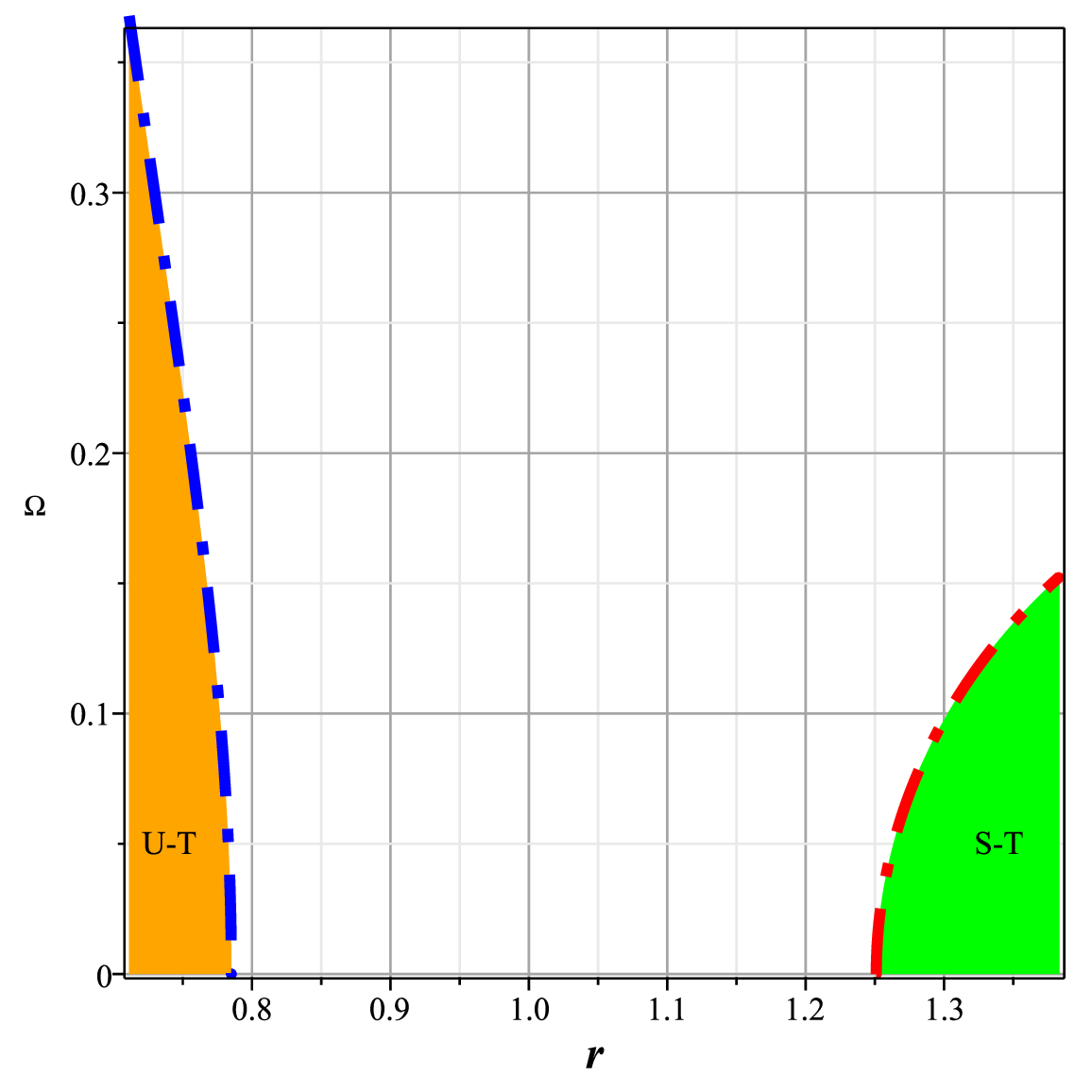}
    \caption{(a) $\beta(r)$ for $M=\ell_0=1$, $\alpha=7.8$, with allowed ($\beta>0$) and forbidden ($\beta<0$) regions. The condition $\beta>0$ is necessary for TCOs but not sufficient; the TCOs also require $\mathcal A'\ge0$ (Fig.~\ref{fig:msco}). (b) $\Omega_{\mathrm{CO}}(r)$ on the branches D1 and D2; between the static spheres $\Omega_{\mathrm{CO}}^{2}<0$. On D2, $\Omega_{\mathrm{CO}}$ increases with $r$ from zero at the stable static sphere.}
    \label{fig:angular_velocity}
\end{figure}

Timelike circular orbits exist only where the conditions~\eqref{eq:tco_domain} hold, namely $\beta=D/2>0$ and $\mathcal A'\ge0$. For $M=\ell_0=1$, $\alpha=7.8$ the first condition holds for $0.7110<r<1.3837$ and for $r>2.5940$ (Fig.~\ref{fig:angular_velocity}a); the interval $r_h<r<0.7110$ lies behind the unstable photon sphere, and $1.3837<r<2.5940$ is bounded by the stable and the outer unstable photon sphere. The second condition fails for $0.7850<r<1.2510$, between the two static spheres, where $\mathcal A'<0$: there $\Omega_{\mathrm{CO}}^{2}<0$ and $\mathcal L_{z,\mathrm{CO}}^{2}<0$, so no circular geodesic exists although $\mathcal E_{\mathrm{CO}}$ is real. The TCO domain thus consists of three disconnected branches (Table~\ref{tab:branches} and Fig.~\ref{fig:msco}).

\begin{table}[htbp]
\centering
\footnotesize
\begin{tabular}{@{} l l l l @{}}
\hline
Branch & $r$-range & Stability & $\Omega_{\mathrm{CO}}$ \\
\hline
D1 & $(0.7110,\,0.7850]$ & unstable & decreases, $0.3685\to0$ \\
D2 & $[1.2510,\,1.3837)$ & stable & increases, $0\to0.1527$ \\
D3a & $(2.5940,\,5.9151)$ & unstable & decreases from $0.2013$ \\
MSCO & $r\simeq5.9151$ & marginal & $0.0694$ \\
D3b & $(5.9151,\,\infty)$ & stable & decreases \\
\hline
\end{tabular}
\caption{Timelike circular orbits for $M=\ell_0=1$, $\alpha=7.8$. D1 runs from the unstable photon sphere to the unstable static sphere, D2 from the stable static sphere to the stable photon sphere, and D3 begins at the outer unstable photon sphere. Photon-sphere endpoints are excluded from the timelike-orbit domain, whereas static-sphere endpoints are included as the zero-angular-momentum limit. The values of $\Omega_{\mathrm{CO}}$ at photon-sphere endpoints are limiting values, and all numerical endpoints are rounded. The MSCO is listed separately from the strictly stable and unstable segments.}
\label{tab:branches}
\end{table}

The zeros of $V''_{\mathrm{eff}}$, Eq.~\eqref{eq:Vpp}, are located at $r=1.1005$, $1.5671$, and $5.9151$. The first two lie outside the TCO domain: at $r=1.1005$ one has $\mathcal A'<0$ and $\mathcal L_{z,\mathrm{CO}}^{2}<0$, and at $r=1.5671$ one has $\mathcal L_{z,\mathrm{CO}}^{2}<0$ and $\mathcal E_{\mathrm{CO}}^{2}<0$. They are formal roots and do not correspond to marginally stable orbits. The only MSCO is $r_{\mathrm{MSCO}}=5.9151$, which separates the unstable branch D3a from the stable branch D3b. On D1 one finds $V''_{\mathrm{eff}}<0$ throughout (it tends to $\mathcal A''(r_{\mathrm{sp1}})<0$ at the unstable static sphere), and on D2 one finds $V''_{\mathrm{eff}}>0$ throughout (it tends to $\mathcal A''(r_{\mathrm{sp2}})>0$ at the stable static sphere). The change from unstable to stable behaviour between the two inner branches therefore does not occur through an MSCO; it takes place across the region $0.7850<r<1.2510$ in which no circular geodesic exists.

Since $\mathrm{d}\Omega_{\mathrm{CO}}^{2}/\mathrm{d}r=(r\mathcal A''-\mathcal A')/(2r^{2})$ on the circular-orbit branches, the sign of $r\mathcal A''-\mathcal A'$ determines the monotonicity of $\Omega_{\mathrm{CO}}$ (Sec.~\ref{sec:origins}). Evaluating it on each branch shows that it is negative on D1 and D3 and positive throughout D2. Hence the angular velocity, Eq.~\eqref{eq:Omega_CO}, decreases with $r$ on D1 (from $0.3685$ at the unstable photon sphere to zero at the unstable static sphere) and on D3, and increases on D2 from zero at the stable static sphere to $0.1527$ at the stable photon sphere $r=1.3837$ (Fig.~\ref{fig:angular_velocity}b). It is therefore monotonic on each connected branch and non-monotonic only across the branches. The increase of $\Omega_{\mathrm{CO}}$ with radius on D2 is the Aschenbach-like behaviour in the sense of Refs.~\cite{7,11}. The local speed $v=r\Omega_{\mathrm{CO}}/\sqrt{\mathcal A}=\sqrt{r\mathcal A'/2\mathcal A}$ measured by static observers increases on D2 from zero and approaches the speed of light as the photon sphere is approached.

\subsection{Static Spheres}
\begin{figure}[htbp]
    \centering
    \includegraphics[width=0.82\linewidth]{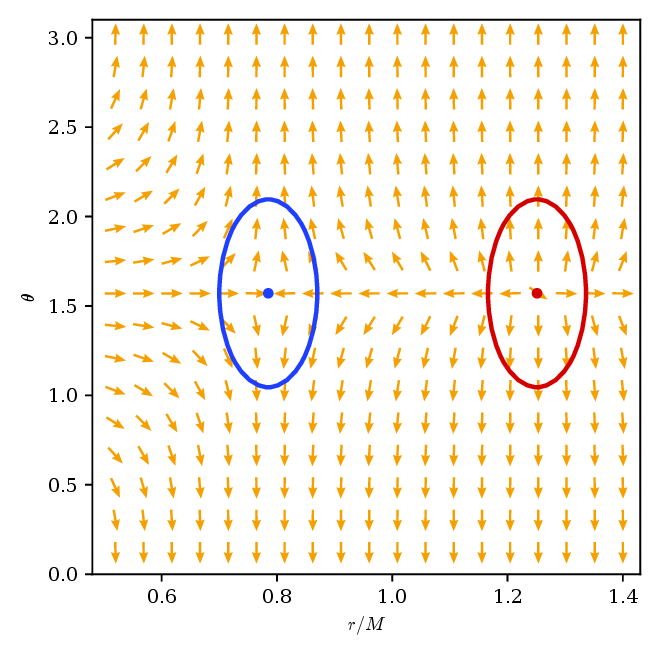}
    \hfill
    \includegraphics[width=0.85\linewidth]{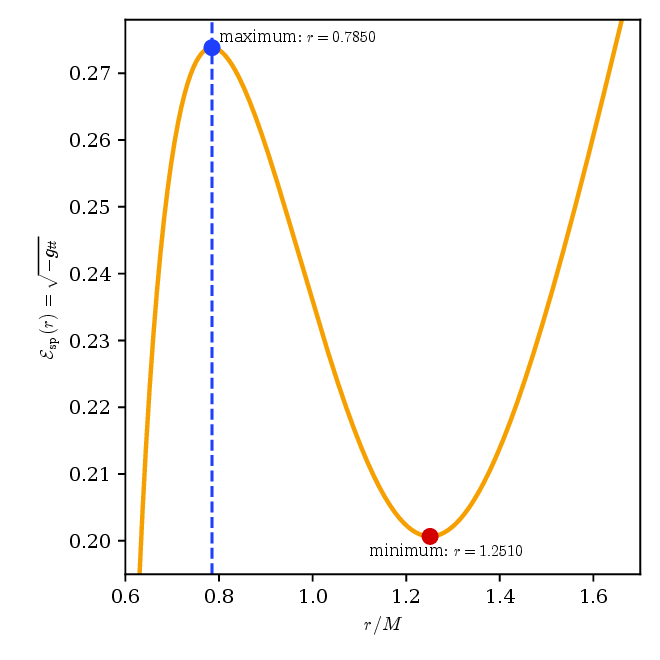}
    \caption{(a) The unstable and stable static spheres at $(r,\theta)=(0.7850,\pi/2)$ and $(1.2510,\pi/2)$ in the $(r,\theta)$ plane for $M=\ell_0=1$, $\alpha=7.8$. (b) $\mathcal E_{\mathrm{sp}}(r)=\sqrt{-g_{tt}}$, whose local maximum (unstable) and minimum (stable) are the static spheres.}
    \label{fig:static_spheres}
\end{figure}

Solving $\mathcal{A}'(r_{\mathrm{sp}}) = 0$ for $M=1, \ell_0=1, \alpha=7.8$ yields two static sphere radii (Fig.~\ref{fig:static_spheres}):
\begin{itemize}
\item Unstable static sphere ($r_{\mathrm{sp1}} = 0.7850$, $\mathcal{E}_{\mathrm{sp}} = 0.2738$, $\mathcal{A}''(r_{\mathrm{sp1}}) = -1.777$),
\item Stable static sphere ($r_{\mathrm{sp2}} = 1.2510$, $\mathcal{E}_{\mathrm{sp}} = 0.2007$, $\mathcal{A}''(r_{\mathrm{sp2}}) = +0.513$).
\end{itemize}
For these parameters $\mathcal A'(r_{\mathrm{sp}})=0$ implies $\mathcal A(r_{\mathrm{sp}})=(1-1/r_{\mathrm{sp}})^{2}$, so that $\mathcal E_{\mathrm{sp}}=|1-1/r_{\mathrm{sp}}|$. The topological charges are $w_1 = -1$ and $w_2 = +1$, giving a total charge $\mathcal{W} = 0$~\cite{13}.

\begin{figure}[htbp]
    \centering
    \includegraphics[width=0.95\linewidth]{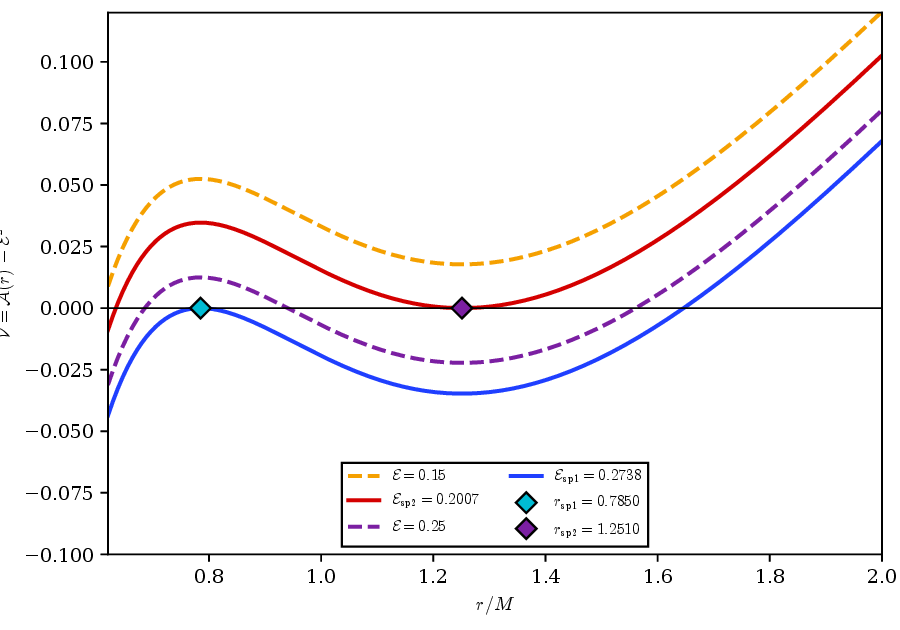}
    \caption{Reduced effective potential $\mathcal V=\mathcal A(r)-\mathcal E^{2}$ of a massive test particle with $\mathcal L_z=0$, for $\mathcal E=0.15$, $0.2007$, $0.25$, and $0.2738$ ($M=\ell_0=1$, $\alpha=7.8$). The curves with $\mathcal E=\mathcal E_{\mathrm{sp1}}$ and $\mathcal E_{\mathrm{sp2}}$ touch zero at the extrema of $\mathcal A$.}
    \label{fig:reduced_potential_energies}
\end{figure}

For the reduced effective potential shown in Fig.~\ref{fig:reduced_potential_energies}, only the curves with $\mathcal{E}=\mathcal{E}_{\mathrm{sp2}}=0.2007$ and $\mathcal{E}=\mathcal{E}_{\mathrm{sp1}}=0.2738$ touch zero, at the local minimum $r_{\mathrm{sp2}}=1.2510$ and at the local maximum $r_{\mathrm{sp1}}=0.7850$ of $\mathcal A$, respectively. The curves with $\mathcal E=0.15$ and $0.25$ do not touch zero at an extremum.

The relation of the static spheres to the SEC follows from Eqs.~\eqref{eq:komar_int} and~\eqref{eq:pt_at_sp}. The unstable static sphere $r_{\mathrm{sp1}}=0.7850$ lies in the region $r<1.0$ where $p_{t,\mathrm{eff}}<0$, and $p_{t,\mathrm{eff}}(r_{\mathrm{sp1}})=\mathcal A''/16\pi=-0.0354$. The stable static sphere $r_{\mathrm{sp2}}=1.2510$ lies in the region $r>1.0$, where the SEC holds, and $p_{t,\mathrm{eff}}(r_{\mathrm{sp2}})=+0.0102$. For both spheres $16\pi\int_{r_h}^{r_{\mathrm{sp}}}r^{2}p_{t,\mathrm{eff}}\,\mathrm{d}r=-r_h^{2}\mathcal A'(r_h)=-0.3478$, as Eq.~\eqref{eq:komar_int} requires (we checked this numerically). The repulsive region $\mathcal A'<0$ ($0.7850<r<1.2510$) extends beyond the SEC-violating region, since $\mathcal A'$ reflects the accumulated effect of the negative $p_{t,\mathrm{eff}}$ at smaller radii.

\section{Parameter-Space Structure and the Origin of the Aschenbach-Like Behaviour}
\label{sec:param}

Section~\ref{sec:hsbh} treats $\ell_0=M$, $\alpha=7.8$ as a representative case. Here we place it in the parameter space of the family. Unless stated otherwise, $0<\ell_0\le M$; the case $\ell_0>M$ is commented on at the end of Sec.~\ref{sec:regimes}. In Secs.~\ref{sec:scaling}--\ref{sec:regimes} the dimensionless radius $x=r/\mathcal M$ is used and derivatives with respect to $x$ are written $\mathcal A_{,x}$, $\mathcal A_{,xx}$; from Sec.~\ref{sec:orbits} on, primes again denote derivatives with respect to $r$.

\subsection{Scaling and horizon}\label{sec:scaling}
Introduce $x=r/\mathcal M$ and $\mu=M/\mathcal M=2M/(2M-\ell_0)$. Then
\begin{equation}
\mathcal A=1-\frac{2\mu}{x}+\alpha e^{-x},
\label{eq:A_scaled}
\end{equation}
so that the family depends on the two dimensionless parameters $(\mu,\alpha)$ only. One has $\ell_0/M=2(1-1/\mu)$, and $\ell_0\le M$ is equivalent to $\mu\le2$. Define
\begin{equation}
\mathcal Q(x)=x^{2}-2\mu x+2\mu .
\end{equation}
For $1<\mu\le2$ the discriminant $4\mu(\mu-2)$ of $\mathcal Q$ is non-positive, so $\mathcal Q\ge0$, with equality only at $x=2$ when $\mu=2$.

\begin{lemma}[Horizon]\label{lem:horizon}
For $1<\mu\le2$ and $\alpha>0$ the function $\mathcal A$ has exactly one zero $x_h$, with $\mathcal A<0$ for $x<x_h$ and $\mathcal A>0$ for $x>x_h$. The zero is simple except at $(\mu,\alpha)=(2,e^{2})$, where it is a triple zero at $x=2$, i.e.\ at $r=M$.
\end{lemma}
\begin{proof}
Write $\mathcal A=e^{-x}[\alpha-\alpha_h(x)]$ with $\alpha_h=(2\mu-x)e^{x}/x$ for $0<x<2\mu$; for $x\ge2\mu$ one has $\mathcal A>0$. Since $\alpha_h'=-e^{x}\mathcal Q/x^{2}\le0$, $\alpha_h$ decreases from $+\infty$ to $0$ on $(0,2\mu)$, which gives the first statement. At the zero, $\mathcal A_{,x}=-e^{-x}\alpha_h'$, which is positive unless $\mathcal Q=0$, i.e.\ unless $(\mu,x)=(2,2)$. For $\mu=2$ one has $\mathcal Q=(x-2)^{2}$ and $\alpha_h(2)=e^{2}$, so $\alpha_h-\alpha$ has a triple zero at $x=2$ for $\alpha=e^{2}$.
\end{proof}

\subsection{Static spheres: exact threshold}\label{sec:static_threshold}
\begin{proposition}[Static spheres]\label{prop:static}
Let $1<\mu\le2$.
\begin{enumerate}
\item[(a)] $\mathcal A_{,x}=0$ if and only if $\alpha=\alpha_{\rm sp}(x)\equiv2\mu e^{x}/x^{2}$. The function $\alpha_{\rm sp}$ has a single minimum, at $x=2$, with value
\begin{equation}
\alpha_{\rm sp}^{*}=\frac{\mu e^{2}}{2}=\frac{e^{2}M}{2M-\ell_0}.
\label{eq:alpha_sp}
\end{equation}
\item[(b)] There is no static sphere for $\alpha<\alpha_{\rm sp}^{*}$. For $\alpha=\alpha_{\rm sp}^{*}$ there is a single degenerate stationary point of $\mathcal A$ at $r=2\mathcal M=2M-\ell_0$, with $\mathcal A_{,x}=\mathcal A_{,xx}=0$ and $\mathcal A=(M-\ell_0)/(2M-\ell_0)$; for $\mu<2$ this is an equilibrium point outside the horizon, whereas for $\mu=2$ it coincides with the triple zero of Lemma~\ref{lem:horizon} and is not an exterior equilibrium. For $\alpha>\alpha_{\rm sp}^{*}$ there are exactly two non-degenerate static spheres, at $x_1<2<x_2$.
\item[(c)] At a static sphere $\mathcal A=\mathcal Q/x^{2}$ and $\mathcal A_{,xx}=2\mu(x-2)/x^{3}$. Hence $x_1$ is a maximum of $\mathcal A$ (unstable) and $x_2$ a minimum (stable).
\item[(d)] The static spheres with $\alpha>\alpha_{\rm sp}^{*}$ lie outside the event horizon.
\end{enumerate}
\end{proposition}
\begin{proof}
(a) and (b) follow from $\mathcal A_{,x}=2\mu/x^{2}-\alpha e^{-x}$ and $\alpha_{\rm sp}'=2\mu e^{x}(x-2)/x^{3}$, together with $\alpha_{\rm sp}\to\infty$ for $x\to0$ and $x\to\infty$. (c) follows by inserting $\alpha e^{-x}=2\mu/x^{2}$ into $\mathcal A$ and into $\mathcal A_{,xx}=-4\mu/x^{3}+\alpha e^{-x}$. (d) By (c), $\mathcal A>0$ at a static sphere unless $\mathcal Q=0$, which can occur only for $\mu=2$ at $x=2$, i.e.\ at $\alpha=e^{2}$; by Lemma~\ref{lem:horizon}, $\mathcal A>0$ means $x>x_h$.
\end{proof}

Since $\mathcal A''=\mathcal A_{,xx}/\mathcal M^{2}$, Eq.~\eqref{eq:pt_at_sp} gives $p_{t,\mathrm{eff}}(r_{\mathrm{sp}})=\mathcal A_{,xx}/(16\pi\mathcal M^{2})$. Proposition~\ref{prop:static}(b) therefore shows that the pair emerges at the radius $r=2M-\ell_0$ where $p_{t,\mathrm{eff}}$ vanishes, so that the combination $\rho_{\mathrm{eff}}+p_{r,\mathrm{eff}}+2p_{t,\mathrm{eff}}$ is saturated there. For $\mu<2$ the second SEC inequality is satisfied at that radius, $\rho_{\mathrm{eff}}+p_{t,\mathrm{eff}}>0$ (Sec.~\ref{sec:sec}), so the SEC is marginally satisfied, and violated only at smaller radii. This is the degenerate static sphere discussed in Ref.~\cite{song}, here in closed form. For $\mu=2$ the critical configuration lies on the horizon.

Since $\mathcal A(x=2)=1-\mu+\alpha e^{-2}$, Lemma~\ref{lem:horizon} implies that the SEC, which is violated for $x<2$, is violated outside the event horizon if and only if
\begin{equation}
\alpha>\alpha_{\rm SEC}\equiv(\mu-1)e^{2}=\frac{e^{2}\ell_0}{2M-\ell_0}.
\end{equation}
Therefore $\alpha_{\rm sp}^{*}/\alpha_{\rm SEC}=M/\ell_0\ge1$: the creation of the static-sphere pair requires a SEC-violating region of finite strength, and the two thresholds coincide, both being equal to $e^{2}$, only for $\ell_0=M$.

\subsection{Photon spheres: exact window}\label{sec:photon_window}
Let $P\equiv2\mathcal A-x\mathcal A_{,x}$; this is the quantity $D$ of Sec.~\ref{sec:tco} in the scaled variable. Photon spheres are the zeros of $P$, and $\mathcal H(x)=\sqrt{\mathcal A}/x$ satisfies $\mathcal H_{,x}=-P/(2x^{2}\sqrt{\mathcal A})$ (the topological potential $\sqrt{\mathcal A}/r$ differs from it by the constant factor $1/\mathcal M$, which does not affect signs). With Eq.~\eqref{eq:A_scaled}, $P=2-6\mu/x+\alpha e^{-x}(x+2)$.

\begin{proposition}[Photon spheres]\label{prop:photon}
Let $1<\mu\le2$ and $\alpha>0$.
\begin{enumerate}
\item[(a)] $P=0$ if and only if $\alpha=\alpha_{\rm ph}(x)\equiv2(3\mu-x)e^{x}/[x(x+2)]$ with $0<x<3\mu$, and
\begin{equation}
\alpha_{\rm ph}'=-\frac{2e^{x}c(x)}{x^{2}(x+2)^{2}},
\label{eq:cubic}
\end{equation}
where $c(x)=x^{3}-(3\mu-1)x^{2}+6\mu$ is the cubic referred to below.
At a photon sphere $\mathcal A=\mathcal Q/[x(x+2)]\ge0$, with equality only at $(\mu,x)=(2,2)$, where the photon sphere coincides with the triple zero of $\mathcal A$ for $\alpha=e^{2}$ and is not exterior. Otherwise photon spheres lie outside the horizon.
\item[(b)] The discriminant of $c$ is $12\mu(3\mu+2)(18\mu^{2}-30\mu-1)$. Let $\mu_c\equiv(5+3\sqrt3)/6$, i.e.\ $\ell_{0c}\equiv(32-18\sqrt3)M\simeq0.8231M$. For $\mu<\mu_c$, $c$ has no positive root, $\alpha_{\rm ph}$ is strictly decreasing, and there is exactly one photon sphere for every $\alpha>0$; it is non-degenerate and unstable. For $\mu=\mu_c$, $c\ge0$ has a double positive root at $x_c=1+\sqrt3$; $\alpha_{\rm ph}$ is decreasing with an inflection point there, there is exactly one photon sphere for every $\alpha>0$, and it is degenerate ($P_{,x}=0$) at $\alpha=\alpha_{\rm cusp}=e^{1+\sqrt3}/(1+\sqrt3)\simeq5.6237$.
\item[(c)] For $\mu>\mu_c$, $c$ has two positive roots $x_a<x_b$, with $x_a<2(3\mu-1)/3<x_b<3\mu$. Let $\alpha_{\rm lo}=\alpha_{\rm ph}(x_a)<\alpha_{\rm hi}=\alpha_{\rm ph}(x_b)$. For $\alpha_{\rm lo}<\alpha<\alpha_{\rm hi}$ there are exactly three non-degenerate photon spheres, unstable, stable, unstable in order of increasing radius. For $\alpha<\alpha_{\rm lo}$ and for $\alpha>\alpha_{\rm hi}$ there is exactly one non-degenerate photon sphere, unstable. For $\alpha=\alpha_{\rm lo}$ and $\alpha=\alpha_{\rm hi}$ there are two distinct photon radii, one non-degenerate and one degenerate. Photon spheres on decreasing branches of $\alpha_{\rm ph}$ are unstable and those on the increasing branch $(x_a,x_b)$ are stable.
\end{enumerate}
\end{proposition}
\begin{proof}
(a) The first statement follows from $P=0$, the second from direct differentiation, and $\mathcal A$ at a photon sphere follows by eliminating $\alpha e^{-x}$ with $P=0$; positivity follows from $\mathcal Q\ge0$ and Lemma~\ref{lem:horizon}. (b), (c) $\alpha_{\rm ph}\to+\infty$ for $x\to0$ and $\alpha_{\rm ph}(3\mu)=0$; $c(0)=6\mu>0$ and $c$ has the sign pattern $(+,-,+)$, hence $0$ or $2$ positive roots counted with multiplicity, and the discriminant is positive exactly for $\mu>\mu_c$ and vanishes at $\mu_c$. For $\mu<\mu_c$, $c>0$ on $x>0$. At $\mu_c$ the minimum of $c$, located at $x_m=2(3\mu-1)/3=1+\sqrt3$, is zero. For $\mu>\mu_c$ the minimum of $c$ at $x_m$ is negative, and $c(3\mu)=9\mu^{2}+6\mu>0$ puts $x_b$ below $3\mu$. On $(0,x_a)$ and $(x_b,3\mu)$ $\alpha_{\rm ph}$ is decreasing, on $(x_a,x_b)$ increasing, which gives the root counts, with the degenerate roots at the extrema $x_a$, $x_b$. Differentiating $P(x;\alpha_{\rm ph}(x))=0$ gives $P_{,x}=-\alpha_{\rm ph}'\,e^{-x}(x+2)$, which has the sign opposite to $\alpha_{\rm ph}'$; thus $\mathcal H_{,xx}=-P_{,x}/(2x^{2}\sqrt{\mathcal A})$ has the sign of $\alpha_{\rm ph}'$, so $\mathcal H$ has a maximum (unstable photon sphere) on decreasing branches and a minimum (stable photon sphere) on the increasing branch.
\end{proof}

\paragraph{Special values.} At $\mu=2$ ($\ell_0=M$) one has $x_a=2$ (that is, $r=M$) and $x_b=(3+\sqrt{33})/2$ (that is, $r=(3+\sqrt{33})M/4\simeq2.186M$), and $\alpha_{\rm lo}=e^{2}$, $\alpha_{\rm hi}\simeq9.2569$. These are the thresholds quoted in Sec.~\ref{sec:hsbh}.

\subsection{Regimes in the $(\ell_0/M,\alpha)$ plane}\label{sec:regimes}
Propositions~\ref{prop:static} and~\ref{prop:photon} give the numbers $(n_{\rm ps},n_{\rm sp})$ of photon spheres and static spheres outside the horizon. The following classification applies to the open, non-degenerate parameter regions; degenerate configurations on the boundary curves are listed below the table. For $1<\mu\le2$, $n_{\rm ps}=3$ if $\mu>\mu_c$ and $\alpha_{\rm lo}<\alpha<\alpha_{\rm hi}$ and $n_{\rm ps}=1$ otherwise, while $n_{\rm sp}=2$ if $\alpha>\alpha_{\rm sp}^{*}$ and $n_{\rm sp}=0$ if $\alpha<\alpha_{\rm sp}^{*}$. This defines four regimes:
\begin{center}
\small
\begin{tabular}{@{} c c c @{}}
\hline
Regime & $n_{\rm ps}$ & $n_{\rm sp}$ \\
\hline
I & 1 & 0 \\
II & 3 & 0 \\
III & 3 & 2 \\
IV & 1 & 2 \\
\hline
\end{tabular}
\end{center}
On the boundary curves the configurations are degenerate: on $\alpha=\alpha_{\rm sp}^{*}$ the static spheres merge into one degenerate stationary point; on $\alpha=\alpha_{\rm lo}$ and $\alpha=\alpha_{\rm hi}$ a pair of photon spheres merges into one degenerate photon sphere; at $\mu=\mu_c$, $\alpha=\alpha_{\rm cusp}$ the single photon sphere is degenerate; and at $(\mu,\alpha)=(2,e^{2})$ all of these coincide with the triple zero of $\mathcal A$ on the horizon.

\begin{figure*}[t]
    \centering
    \includegraphics[width=0.92\textwidth]{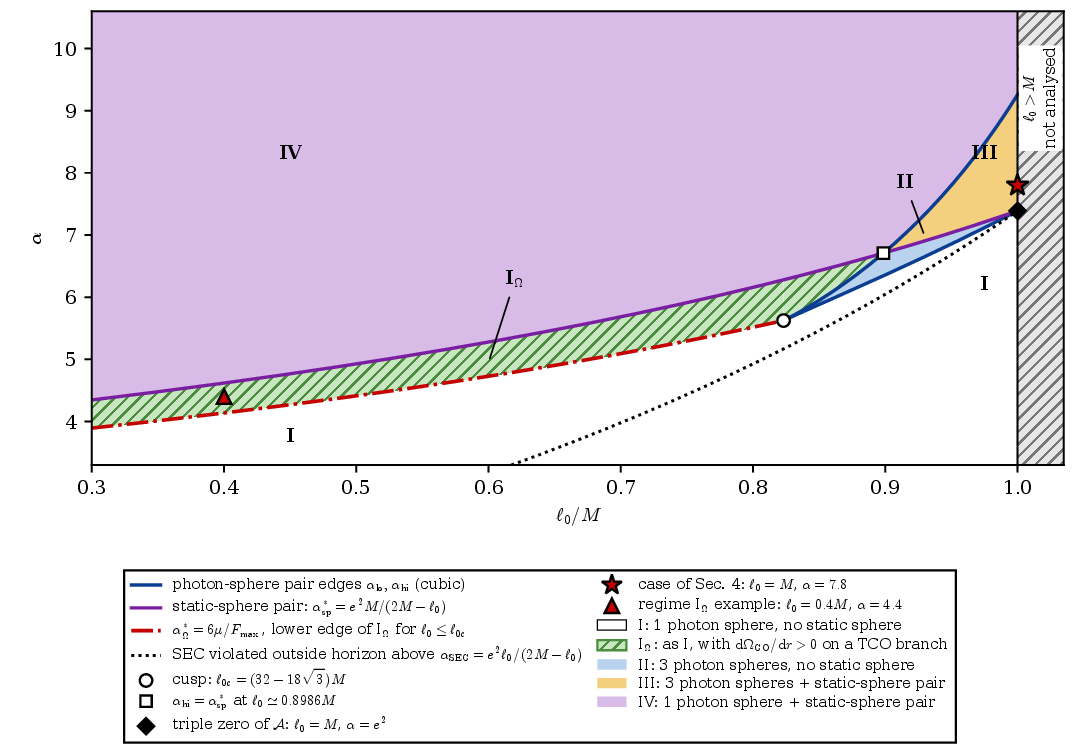}
    \caption{Regimes of the EGD hairy Schwarzschild family in the $(\ell_0/M,\alpha)$ plane for $\ell_0\le M$. I: one photon sphere, no static sphere; II: three photon spheres, no static sphere; III: three photon spheres and a pair of static spheres; IV: one photon sphere and a pair of static spheres. The hatched subregion I$_\Omega$ of regime~I is the part of the parameter space with a circular-orbit branch on which $\Omega_{\rm CO}$ increases although there is neither a static sphere nor a stable photon sphere (Sec.~\ref{sec:origins}). The blue curves are the edges $\alpha_{\rm lo}$ and $\alpha_{\rm hi}$ of the three-photon-sphere window from the cubic~\eqref{eq:cubic}, the purple curve is $\alpha_{\rm sp}^{*}$ of Eq.~\eqref{eq:alpha_sp}, the red dash-dotted curve is $\alpha_\Omega^{*}$ of Proposition~\ref{prop:threshold}, and the dotted curve is $\alpha_{\rm SEC}$. Above the lower boundary formed by the red curve and the blue $\alpha_{\rm lo}$ curve, $\Omega_{\rm CO}$ increases with $r$ on a branch of circular orbits. The star marks the case of Sec.~\ref{sec:hsbh}, the triangle the example of Fig.~\ref{fig:omega_regimes}(a).}
    \label{fig:regime_map}
\end{figure*}

The boundaries are shown in Fig.~\ref{fig:regime_map}. The following ordering was established numerically on grids of $2\times10^{3}$ (item i), $4\times10^{3}$ (item ii) and $2\times10^{4}$ (item iii) values of $\mu\in(\mu_c,2]$: (i) $\alpha_{\rm lo}<\alpha_{\rm sp}^{*}$, with equality at $\mu=2$; (ii) $\alpha_{\rm hi}>\alpha_{\rm sp}^{*}$ if and only if $\mu>\mu_*\simeq1.8159$ (the root of $\alpha_{\rm hi}=\alpha_{\rm sp}^{*}$), i.e.\ $\ell_0>\ell_{0*}\simeq0.8986M$; (iii) $\alpha_{\rm lo}\ge\alpha_{\rm SEC}$, with equality at $\mu=2$. Consequently, regime~II exists for $\ell_{0c}<\ell_0<M$ and regime~III for $\ell_{0*}<\ell_0\le M$. For $\ell_0=M$ the lower edges $\alpha_{\rm lo}=\alpha_{\rm sp}^{*}=\alpha_{\rm SEC}=e^{2}$ coincide, regime~II shrinks to nothing, and the case of Sec.~\ref{sec:hsbh}, $\alpha=7.8$, lies in regime~III. For $\mu>\mu_c$ the single photon sphere is the outer one ($x>x_b$) for $\alpha<\alpha_{\rm lo}$ and the inner one ($x<x_a$) for $\alpha>\alpha_{\rm hi}$; by (i), regime~IV lies in the latter range. Item (iii) shows that, within this family and for $\ell_0\le M$, a stable photon sphere outside the horizon requires a SEC violation outside the horizon, in line with the result of Ref.~\cite{3_2} that SEC-respecting solutions have a single photon sphere, and a stable photon sphere requires $\ell_0>\ell_{0c}$.

\paragraph{Numerical method.} Root counts were obtained by locating sign changes of $\mathcal A$, $P$, and $\mathcal A_{,x}$ on a fine radial grid and refining the roots by bracketing; photon spheres and static spheres were counted outside the largest zero of $\mathcal A$. The assignment of the regimes from Props.~\ref{prop:static} and~\ref{prop:photon} agrees with these counts at $1500$ random points with $1.05\le\mu\le2$ and $0.3\le\alpha\le16$, and at $30$ points adjacent to the boundaries.

\begin{remark}[The case $\ell_0>M$]
For $\mu>2$ the quadratic $\mathcal Q$ changes sign, $\mathcal A$ can have more than one zero, and static or photon spheres located where $\mathcal A<0$ are not outside the event horizon, so Propositions~\ref{prop:static} and~\ref{prop:photon} do not by themselves locate the spheres outside it. We therefore only quote numerical results, obtained by taking the event horizon as the largest zero of $\mathcal A$. For $\ell_0=1.1M$ the static-sphere pair and the additional photon spheres appear together at $\alpha\simeq9.68$, and the three-photon-sphere window extends to $\alpha\simeq14.21$; for $\ell_0=1.2M$ the corresponding values are $14.23$ and $25.27$. The creation thresholds are therefore larger than $\alpha_{\rm sp}^{*}=8.21$ and $9.24$, respectively.
\end{remark}

\subsection{Circular-orbit branches, stability, and the $(\mathcal E,\mathcal L_z)$ diagram}\label{sec:orbits}
\begin{proposition}[Circular orbits for $g_{tt}g_{rr}=-1$]\label{prop:orbits}
Let $g(r)=\mathcal A\mathcal A''-2\mathcal A'^{2}+3\mathcal A\mathcal A'/r$ and $D=2\mathcal A-r\mathcal A'$. On the timelike circular orbits ($D>0$, $\mathcal A'\ge0$):
\begin{enumerate}
\item[(a)] $V''_{\mathrm{eff}}=2g/D$ and $\mathrm d(\mathcal L_{z,\mathrm{CO}}^{2})/\mathrm dr=2r^{3}g/D^{2}$. An orbit is stable if and only if $g>0$, i.e.\ if and only if $\mathrm d\mathcal L_{z,\mathrm{CO}}/\mathrm dr>0$. Non-degenerate zeros of $g$ are MSCOs and correspond to extrema of $\mathcal L_{z,\mathrm{CO}}(r)$; at a double zero $\mathrm d\mathcal L_{z,\mathrm{CO}}/\mathrm dr$ vanishes without changing sign.
\item[(b)] $\mathrm d\mathcal E_{\mathrm{CO}}=\Omega_{\mathrm{CO}}\,\mathrm d\mathcal L_{z,\mathrm{CO}}$ along the family of circular orbits.
\item[(c)] The radial epicyclic frequency with respect to the coordinate time is $\kappa_r^{2}=g/2$. For $\mathcal A=1-2M/r$ this is $\Omega^{2}(1-6M/r)$.
\end{enumerate}
For asymptotically flat metrics $g\to2M/r^{3}>0$ at large $r$.
\end{proposition}
\begin{proof}
(a) and (c) follow by direct differentiation of Eq.~\eqref{eq:E_L_CO}, using $\kappa_r^{2}=(V''_{\mathrm{eff}}/2)(\mathcal A/\mathcal E_{\mathrm{CO}})^{2}$. (b) From $\mathcal V_{\mathrm{eff}}^{(\mathrm{TL})}=0$ one gets $-2\mathcal E\,\mathrm d\mathcal E+2\mathcal A\mathcal L_z\,\mathrm d\mathcal L_z/r^{2}=0$, i.e.\ $\mathrm d\mathcal E/\mathrm d\mathcal L_z=\mathcal A\mathcal L_z/(r^{2}\mathcal E)=\Omega$.
\end{proof}

\begin{proposition}[Static spheres and photon spheres bound the same branches]\label{prop:bridge}
Let $\mathcal A$ be the metric function of a static, asymptotically flat spacetime with $g_{tt}g_{rr}=-1$ and a regular non-extremal horizon, let $\mathcal A'$ have exactly two zeros outside the horizon, the static spheres, and let the zeros of $D$ be simple. Then: (a) the largest zero $r_a$ of $D$ below the inner static sphere $r_{\mathrm{sp1}}$ exists and is an unstable photon sphere, and $(r_a,r_{\mathrm{sp1}}]$ is a branch of timelike circular orbits; (b) the circular orbits that start at the outer static sphere $r_{\mathrm{sp2}}$ form a branch $[r_{\mathrm{sp2}},r_b)$, where $r_b$ is the smallest zero of $D$ above $r_{\mathrm{sp2}}$, a stable photon sphere, or $r_b=\infty$ if there is none.
\end{proposition}
\begin{proof}
Let $\mathcal H(r)=\sqrt{\mathcal A}/r$, so that
$\mathcal H'=-D/(2r^{2}\sqrt{\mathcal A})$; at a simple zero of $D$ one has
$\mathcal H''=-D'/(2r^{2}\sqrt{\mathcal A})$, and $\mathcal H''<0$ (maximum, unstable photon sphere) if and only if $D'>0$.
At the horizon $D=-r_h\mathcal A'(r_h)<0$, and at a static sphere $D=2\mathcal A>0$, so $D$ has a zero in $(r_h,r_{\mathrm{sp1}})$; take the largest one, $r_a$. There $D$ crosses from negative to positive, so $D'>0$ and $\mathcal H''<0$ (unstable). Since $r_{\mathrm{sp1}}$ is the first zero of $\mathcal A'$ outside the horizon, $\mathcal A'>0$ on $(r_h,r_{\mathrm{sp1}})$ and $D>0$ on $(r_a,r_{\mathrm{sp1}}]$, so these are timelike circular orbits. For (b), $\mathcal A'>0$ for $r>r_{\mathrm{sp2}}$ because the two static spheres are the only zeros of $\mathcal A'$, and $D(r_{\mathrm{sp2}})>0$; at $r_b$ $D$ crosses from positive to negative, so $D'<0$ and $\mathcal H''>0$ (stable). For asymptotically flat spacetimes $D\to2$ at infinity.
\end{proof}
The assumption on $\mathcal A'$ holds for the family considered here for $\alpha>\alpha_{\rm sp}^{*}$ (Proposition~\ref{prop:static}).

\begin{figure}[t]
    \centering
    \includegraphics[width=\linewidth]{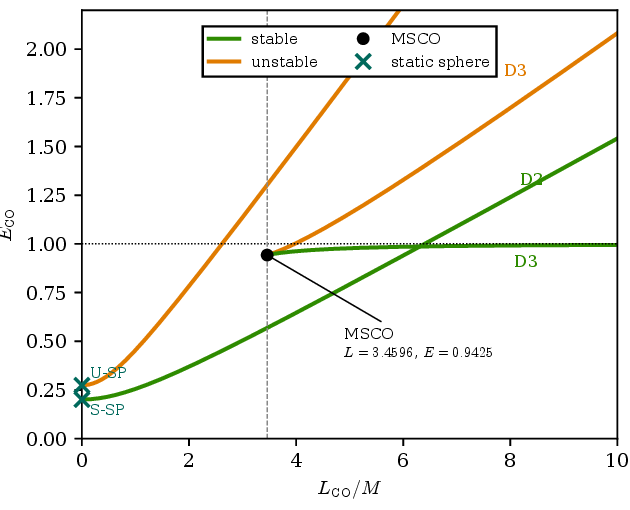}
    \caption{Energy $\mathcal E_{\rm CO}$ versus angular momentum $\mathcal L_{\rm CO}$ of the timelike circular orbits for $M=\ell_0=1$, $\alpha=7.8$. D1 and D3 (upper branch) are unstable, D2 and D3 (lower branch) are stable; the static spheres lie at $\mathcal L=0$ and the three photon spheres at $\mathcal L,\mathcal E\to\infty$. The slope of each curve is $\Omega_{\rm CO}$. The MSCO is the cusp at $\mathcal L=3.4596$, $\mathcal E=0.9425$.}
    \label{fig:EL}
\end{figure}

For the representative case $\ell_0=M$, $\alpha=7.8$ the sign of $g$ is negative on the whole branch D1 and positive on the whole branch D2 (numerically), so by Proposition~\ref{prop:orbits}(a) $\mathcal L_{z,\mathrm{CO}}$ decreases from $\infty$ at the unstable photon sphere to $0$ at the unstable static sphere along D1, and increases from $0$ at the stable static sphere to $\infty$ at the stable photon sphere along D2, while along D3 it decreases from $\infty$ to its minimum $\mathcal L_{\rm MSCO}=3.4596M$ at $r=5.9151M$ and then increases again. Consequently there are exactly two circular orbits for $\mathcal L_z<\mathcal L_{\rm MSCO}$, one unstable on D1 and one stable on D2, three for $\mathcal L_z=\mathcal L_{\rm MSCO}$ (the two inner ones and the marginal orbit), and exactly four for $\mathcal L_z>\mathcal L_{\rm MSCO}$ (Fig.~\ref{fig:EL}). The inner pair is the continuation to $\mathcal L_z\neq0$ of the pair of static spheres. In this solution each static sphere is joined, through a branch of circular orbits, to a photon sphere of the same topological charge, $-1$ for the unstable and $+1$ for the stable one. The outer pair has $\mathcal E_{\rm MSCO}=0.94254$ and efficiency $1-\mathcal E_{\rm MSCO}=5.746\%$, against $\mathcal E=0.94281$, $\mathcal L_z=3.4641M$, $r=6M$ and $5.719\%$ for Schwarzschild; the deformation shifts the outer MSCO by $1.4\%$ but the outer unstable photon sphere ($2.5940M$) by $13.5\%$ with respect to Schwarzschild ($3M$). For comparison, Ref.~\cite{11} reports a single pair of circular orbits per angular momentum for the massive-gravity black holes it considers, and two pairs for the dyonic quasi-topological electromagnetism black hole~\cite{7,11}. The EGD family contains both behaviours: two pairs above $\mathcal L_{\rm MSCO}$ in the representative case, and a single pair for every $\mathcal L_z$ in the examples without an MSCO discussed below.

\subsection{Where the angular velocity increases}
\label{sec:origins}
Along a branch of circular orbits ($\mathcal A'>0$), $\mathrm d\Omega_{\mathrm{CO}}^{2}/\mathrm dr=(r\mathcal A''-\mathcal A')/(2r^{2})$, so that
\begin{equation}
\frac{\mathrm d\Omega_{\mathrm{CO}}}{\mathrm dr}>0\iff r\mathcal A''>\mathcal A' .
\label{eq:criterion}
\end{equation}
This criterion and several of its consequences were obtained in Ref.~\cite{7}; we restate those used below.

\begin{lemma}[Consequences of Eq.~\eqref{eq:criterion}~\cite{7}]\label{lem:criterion}
\begin{enumerate}
\item[(a)] $r\,g=2\mathcal A'D+\mathcal A(r\mathcal A''-\mathcal A')$. Hence on circular orbits an increasing $\Omega_{\mathrm{CO}}$ implies $g>0$: increasing segments lie on stable branches.
\item[(b)] At a non-degenerate photon sphere $\mathrm d\Omega_{\mathrm{CO}}^{2}/\mathrm dr=-D'/(2r^{2})$, which equals $\sqrt{\mathcal A}\,\mathcal H''$ for $\mathcal H=\sqrt{\mathcal A}/r$: $\Omega_{\mathrm{CO}}$ increases at a stable photon sphere and decreases at an unstable one.
\item[(c)] Near a stable static sphere $\Omega_{\mathrm{CO}}\simeq[\mathcal A''(r_{\mathrm{sp}})(r-r_{\mathrm{sp}})/(2r_{\mathrm{sp}})]^{1/2}$ for $r>r_{\mathrm{sp}}$, so $\Omega_{\mathrm{CO}}$ starts at zero with $\mathrm d\Omega_{\mathrm{CO}}/\mathrm dr\to+\infty$. Near the unstable static sphere $\Omega_{\mathrm{CO}}\to0$ with $\mathrm d\Omega_{\mathrm{CO}}/\mathrm dr\to-\infty$ as $r\to r_{\mathrm{sp1}}^{-}$.
\end{enumerate}
\end{lemma}
\begin{proof}
(a) follows from the definitions of $g$ and $D$ by expanding $2\mathcal A'D+\mathcal A(r\mathcal A''-\mathcal A')$. 
(b) With $\mathcal H'=-D/(2r^{2}\sqrt{\mathcal A})$ and $D=0$ at the photon sphere, $\mathcal H''=-D'/(2r^{2}\sqrt{\mathcal A})$, and $D'=\mathcal A'-r\mathcal A''$. 
(c) Expand $\mathcal A'\simeq\mathcal A''(r_{\mathrm{sp}})(r-r_{\mathrm{sp}})$ in $\Omega^{2}=\mathcal A'/(2r)$.
\end{proof}

Increasing segments are thus guaranteed next to every stable photon sphere and next to every stable static sphere. For the effective source the criterion takes the following form.

\begin{proposition}[Relation to the effective source]\label{prop:Omega_pt}
For $g_{tt}g_{rr}=-1$, with $p_t$ defined by Eq.~\eqref{eq:komar},
\begin{equation}
\frac{\mathrm d\Omega_{\mathrm{CO}}^{2}}{\mathrm dr}=\frac{16\pi r\,p_t-3\mathcal A'}{2r^{2}} .
\label{eq:Omega_pt}
\end{equation}
Consequently, on the interior of a branch of circular orbits ($\mathcal A'>0$), an increasing angular velocity requires $p_t>0$. This is the first of the two SEC inequalities; the second, $\rho+p_t\ge0$, does not follow from it in general. For the exponential source of Sec.~\ref{sec:hsbh}, $p_t>0$ implies $r>2\mathcal M$, where $\rho_{\mathrm{eff}}+p_{t,\mathrm{eff}}\ge0$ also holds, so that the full SEC holds on every increasing segment.
\end{proposition}
\begin{proof}
$\mathrm d\Omega^{2}/\mathrm dr=(r\mathcal A''-\mathcal A')/(2r^{2})$ and, from Eq.~\eqref{eq:komar}, $r\mathcal A''=16\pi r\,p_t-2\mathcal A'$. The last statement follows from Sec.~\ref{sec:sec}: $p_{t,\mathrm{eff}}>0$ iff $r>2\mathcal M$, and $\rho_{\mathrm{eff}}+p_{t,\mathrm{eff}}\ge0$ for $r\ge\sqrt2\,\mathcal M<2\mathcal M$.
\end{proof}
In the representative case the increasing segment $1.2510<r<1.3837$ lies in the region $r>2M-\ell_0=1$, whereas the decreasing branch D1 lies in the SEC-violating region. The SEC violation is thus needed to create the pair of static spheres, but the increasing segment itself lies where $p_t>0$.

Criterion~\eqref{eq:criterion} is local and does not refer to either kind of sphere. For the exponential family it reads $x\mathcal A_{,xx}-\mathcal A_{,x}=-6\mu/x^{2}+\alpha e^{-x}(x+1)$.

\begin{proposition}[Increasing segments in the exponential family]\label{prop:threshold}
Let $F(x)=x^{2}(x+1)e^{-x}$. Then $\mathrm d\Omega_{\mathrm{CO}}^{2}/\mathrm dx>0$ if and only if $\alpha F(x)>6\mu$. $F$ increases on $(0,x_\Omega)$ and decreases beyond, with $x_\Omega=1+\sqrt3$ and $F_{\max}=(1+\sqrt3)^{2}(2+\sqrt3)e^{-(1+\sqrt3)}\simeq1.81305$. Hence the criterion is met somewhere if and only if
\begin{equation}
\alpha>\alpha_\Omega^{*}(\mu)\equiv\frac{6\mu}{F_{\max}}\simeq3.3093\,\mu,
\label{eq:alpha_Omega}
\end{equation}
and then on a single interval $x_-<x<x_+$ around $x_\Omega$. The ratio $\alpha_\Omega^{*}/\alpha_{\rm sp}^{*}=12/(e^{2}F_{\max})\simeq0.896$ does not depend on $\mu$, and $\alpha_\Omega^{*}(\mu_c)=\alpha_{\rm cusp}$.
\end{proposition}
\begin{proof}
$F'=-x(x^{2}-2x-2)e^{-x}$ vanishes at $x_\Omega=1+\sqrt3$. For the last statement, $6\mu_c=x_\Omega(x_\Omega+1)=5+3\sqrt3$, so that $\alpha_\Omega^{*}(\mu_c)=e^{x_\Omega}/x_\Omega=\alpha_{\rm cusp}$.
\end{proof}
The interval $(x_-,x_+)$ gives a physical increasing segment only where it lies outside the horizon and on the circular-orbit domain, $D>0$ and $\mathcal A'>0$. For $\ell_0\le M$ we find numerically that the parameters admitting such a segment are those with $\alpha>\alpha_A(\ell_0)$, where $\alpha_A=\alpha_\Omega^{*}$ for $\ell_0\le\ell_{0c}$ and $\alpha_A=\alpha_{\rm lo}$ for $\ell_{0c}<\ell_0\le M$; the two curves join at the cusp. The numerically determined threshold agrees with this formula to $2\times10^{-6}$ at $21$ values of $\ell_0$ between $0.60M$ and $M$, and a segment is present for all larger $\alpha$ tested, up to $\alpha=40$. For $\ell_0>\ell_{0c}$ the effect therefore starts exactly when the stable photon sphere is created, whereas for smaller $\ell_0$ it starts earlier, at $\alpha_\Omega^{*}$, with neither a stable photon sphere nor a static sphere present. The subregion of regime~I with $\alpha>\alpha_A$ is denoted I$_\Omega$ (Fig.~\ref{fig:regime_map}).

\begin{proposition}[Profiles of $\Omega_{\mathrm{CO}}$ in the regimes]\label{prop:regimes_omega}
Let $1<\mu\le2$.
\begin{enumerate}
\item[(II)] In regime II the branch of circular orbits between the inner unstable photon sphere and the stable photon sphere is connected, and $\Omega_{\mathrm{CO}}$ has an interior minimum on it.
\item[(III)] In regime III, $\Omega_{\mathrm{CO}}$ increases at both ends of the branch D2 that joins the stable static sphere to the stable photon sphere.
\item[(IV)] In regime IV there is a single branch $[r_{\mathrm{sp2}},\infty)$ starting at the stable static sphere, and $\Omega_{\mathrm{CO}}$ has an interior maximum on it. If all zeros of $g$ on this branch are simple, their number is even.
\item[(I$_\Omega$)] In the subregion I$_\Omega$ of regime I there is neither a static sphere nor a stable photon sphere, but $\Omega_{\mathrm{CO}}$ has an interior minimum followed by an interior maximum on the single branch of circular orbits.
\end{enumerate}
\end{proposition}
\begin{proof}
(II) There is no static sphere, so $\mathcal A'>0$ outside the horizon, and $D>0$ between the first two photon spheres (Proposition~\ref{prop:photon}), so the branch is connected. By Lemma~\ref{lem:criterion}(b), $\mathrm d\Omega/\mathrm dr<0$ at the unstable end and $>0$ at the stable end, so $\Omega$ has an interior minimum. (III) By Proposition~\ref{prop:bridge}(b) the branch D2 ends at the stable photon sphere, where $\mathrm d\Omega/\mathrm dr>0$ by Lemma~\ref{lem:criterion}(b), and it starts with $\mathrm d\Omega/\mathrm dr=+\infty$ by Lemma~\ref{lem:criterion}(c). (IV) Here $n_{\rm ps}=1$ and, by Proposition~\ref{prop:bridge}(a), the single photon sphere lies below $r_{\mathrm{sp1}}$; hence $D>0$ for all larger radii, $\mathcal A'>0$ for $r>r_{\mathrm{sp2}}$, and the branch extends to infinity. $\Omega_{\mathrm{CO}}$ vanishes at $r_{\mathrm{sp2}}$, is positive after it, and tends to zero at infinity, so it has an interior maximum. Finally $g(r_{\mathrm{sp2}})=\mathcal A\mathcal A''>0$ and $g\to2M/r^{3}>0$ at infinity, so $g$ has an even number of simple sign changes. (I$_\Omega$) The branch runs from the single, non-degenerate, unstable photon sphere to infinity with $\mathcal A'>0$ and $D>0$. By Proposition~\ref{prop:threshold} the criterion~\eqref{eq:criterion} holds exactly on an interval $(x_-,x_+)$ and fails elsewhere. It fails at the unstable photon sphere (Lemma~\ref{lem:criterion}(b)); since a physical increasing segment exists in I$_\Omega$ (the numerical result above), the interval lies entirely above the photon sphere. Hence $\Omega_{\mathrm{CO}}$ decreases up to $x_-$, increases on $(x_-,x_+)$, and decreases beyond $x_+$, which gives the minimum and the maximum.
\end{proof}
In regime~III we find numerically that $\Omega_{\mathrm{CO}}$ is monotonic on D2 (Sec.~\ref{sec:tco}). At the merger of two MSCOs the zeros of $g$ coalesce into one double zero; for $\ell_0=M$ this occurs, numerically, at $\alpha\simeq30.621$, $r\simeq5.1464M$, and for larger $\alpha$ the branch of regime~IV is stable throughout.

\begin{figure*}[t]
    \centering
    \includegraphics[width=0.78\textwidth]{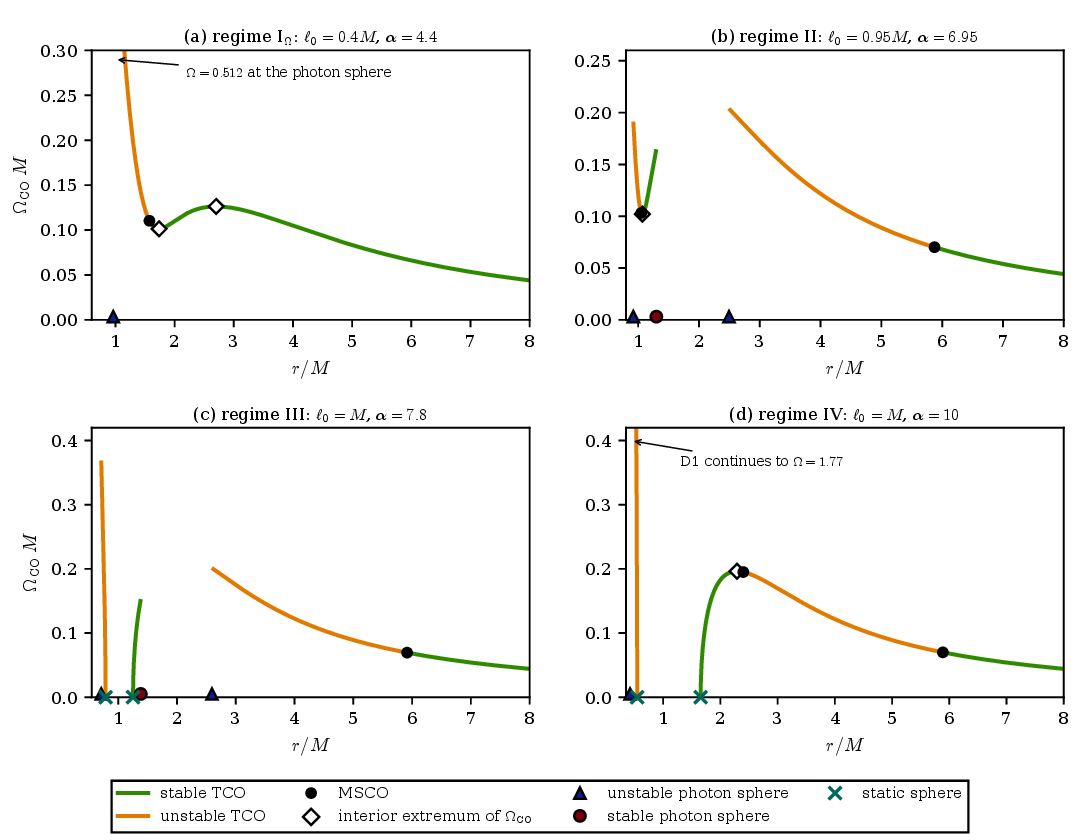}
    \caption{Angular velocity $\Omega_{\rm CO}$ of the timelike circular orbits in (a) regime I$_\Omega$, (b) regime II, (c) regime III, and (d) regime IV ($M=1$). Green: stable, orange: unstable circular orbits; black dots: MSCOs; white diamonds: interior extrema of $\Omega_{\rm CO}$; triangles and circles on the axis: unstable and stable photon spheres; crosses: static spheres. In (a) the branch starts at the photon sphere with $\Omega_{\rm CO}=0.512$ and is cut off; in (d) the inner branch D1 reaches $\Omega=1.77$ at the photon sphere and is cut off.}
    \label{fig:omega_regimes}
\end{figure*}

Examples (computed with $M=1$) are shown in Fig.~\ref{fig:omega_regimes}.
\begin{itemize}
\item Regime I$_\Omega$, $\ell_0=0.4M$, $\alpha=4.4$: horizon at $r_h=0.7133$, a single photon sphere at $r=0.9632$ ($\Omega_{\rm CO}=0.5124$ there), no static sphere. $\Omega_{\rm CO}$ has a minimum $0.1013$ at $r=1.7364$ and a maximum $0.1263$ at $r=2.7018$, so that $\mathrm d\Omega_{\rm CO}/\mathrm dr>0$ on $1.7364<r<2.7018$, where the orbits are stable. For example, at $r=2$: $\mathcal A=0.3612$, $\mathcal A'=0.04853$, $D=0.6253$, $g=0.04482$.
\item Regime II, $\ell_0=0.95M$, $\alpha=6.95$: no static sphere, photon spheres at $r=0.9209$, $1.2993$, $2.4942$, so that $\Omega_{\rm CO}$ decreases from $0.1912$ at the inner photon sphere to a minimum $0.1020$ at $r=1.0704$ and then increases to $0.1646$ at the stable photon sphere. The branch contains an MSCO at $r=1.0521$ ($\mathcal E=0.2311$, $\mathcal L_z=0.7358$), and a second MSCO exists at $r=5.874$.
\item Regime III, $\ell_0=M$, $\alpha=7.8$: the case of Sec.~\ref{sec:hsbh}.
\item Regime IV, $\ell_0=M$, $\alpha=10$: photon sphere at $r=0.4214$, static spheres at $0.5458$ and $1.6552$, a maximum $\Omega_{\rm CO}=0.1962$ at $r=2.2903$, and two MSCOs, at $r=2.4007$ ($\mathcal E=1.4419$, $\mathcal L_z=6.5072$) and $r=5.8872$ ($\mathcal E=0.9425$, $\mathcal L_z=3.4582$). The stable circular orbits adjacent to the inner MSCO have $\mathcal E>1$. For $\ell_0=0.8M$, $\alpha=8$, $\Omega_{\rm CO}$ has its maximum $0.1529$ at $r=2.6835$ and the MSCOs are at $4.1978$ and $5.1437$; for $\ell_0=0.4M$, $\alpha=6$ the maximum is $0.0993$ at $r=3.5780$ and there is no MSCO, so that the branch is stable from $r_{\mathrm{sp2}}$ to infinity and there is a single pair of circular orbits for every $\mathcal L_z$.
\end{itemize}

Thus the increase of $\Omega_{\mathrm{CO}}$ with radius is governed by the local criterion~\eqref{eq:criterion}. It is guaranteed next to every stable photon sphere and next to every stable static sphere~\cite{7}, which gives the profiles of regimes~II, III, and IV, and it also occurs in regime~I$_\Omega$, where neither is present. The profile has an interior minimum in regime~II, an interior minimum followed by an interior maximum in regime~I$_\Omega$, and an interior maximum in regime~IV.

\section{Discussion}

A comparison between different black hole models admitting static spheres and Aschenbach-like profiles is summarized in Table~\ref{tab:comparison}. In non-linear electrodynamics (NLE) and massive gravity, repulsive gravitational effects originate from non-linear field couplings or graviton mass terms~\cite{7,11,13}. In the EGD hairy Schwarzschild metric, the deformation parameter $\alpha$ alters the spacetime geometry, generating an effective stress-energy contribution $T^{\rm eff}_{\mu\nu}=\alpha\theta_{\mu\nu}$.

Our explicit evaluation shows that this effective source violates the SEC for $r < 2M - \ell_0$. For $M=\ell_0=1$ this region lies outside the horizon only for $\alpha>e^{2}$, which is the same threshold at which the static spheres and the additional photon spheres appear. The EGD solution thus illustrates how geometric deformations mimic repulsive matter sources. The inner static sphere lies inside the SEC-violating region; the outer one lies where $p_{t,\mathrm{eff}}>0$ and the SEC holds locally, so that its existence is consistent with the integral relation~\eqref{eq:komar_int}, which requires negative tangential pressure only somewhere between the horizon and the equilibrium radius. For the case of Sec.~\ref{sec:hsbh}, Appendix~B gives the breakdown against the null, weak, and dominant energy conditions as well: only a narrow shell adjoining the horizon violates the NEC, and both the innermost static sphere and the innermost photon sphere lie outside this shell.

Section~\ref{sec:param} shows what the exponential deformation contributes beyond these general results. (i) The solution is neutral and exactly asymptotically flat, its seed is the vacuum Schwarzschild solution, and the deformation has the finite range $\mathcal M$; in the massive-gravity examples of Ref.~\cite{11} the lapse function contains a $\Lambda r^{2}/3$ term with $\Lambda\neq0$, and the dyonic solutions carry electromagnetic charges. (ii) Because the deformation is a single exponential, the whole family depends on the two parameters $(\mu,\alpha)$, and the creation of the static-sphere pair and of the additional photon spheres is controlled by closed-form conditions (Propositions~\ref{prop:static} and~\ref{prop:photon}); the pair is created exactly on the boundary $r=2M-\ell_0$ of the SEC-violating region. (iii) The increase of the angular velocity of circular orbits is governed by the local criterion $r\mathcal A''>\mathcal A'$~\cite{7}. It is guaranteed next to every stable photon sphere and every stable static sphere; regimes II and IV isolate the first and the second case, regime III combines them, and in the part I$_\Omega$ of regime I the increase occurs with neither present. These statements concern the test-particle dynamics. The absence of an electromagnetic or other matter field in the construction does not make the effective source benign: the DEC is violated far from the black hole, for $r>(2+\sqrt2)\mathcal M$, for all parameters, and the NEC is violated in a shell adjoining the horizon whenever $r_h<\sqrt2\,\mathcal M$ (Appendix~B). The latter holds for the parameters of Sec.~\ref{sec:hsbh} but not throughout the family; for example $r_h=0.7529M>\sqrt2\,\mathcal M=0.7425M$ for $\ell_0=0.95M$, $\alpha=6.95$. The fundamental origin of the source is not specified. The comparison with Refs.~\cite{7,11,13} is therefore one of geometric structure and not of physical realisability.

\begin{table*}[htbp]
\centering
\small
\begin{tabular}{@{} p{3.2cm} p{3.7cm} p{3.2cm} p{5.5cm} @{}}
\hline
\textbf{Property / Model} & \textbf{NLE / Dyonic BHs~\cite{5_2,13}} & \textbf{Massive Gravity~\cite{11}} & \textbf{EGD Hairy (This Work)} \\
\hline
Effective Repulsion Origin & Non-linear EM fields & Graviton mass terms & Geometric deformation, effective source $T^{\rm eff}_{\mu\nu}=\alpha\theta_{\mu\nu}$ \\
Matter content and asymptotics & Charged, asymptotically flat & Examples with $\Lambda\neq0$~\cite{11} & Neutral, asymptotically flat, finite-range deformation \\
SEC Status ($r > r_h$) & Local violation & Local violation & Local violation for $r < 2M - \ell_0$ \\
Stable Photon Sphere & Present & Present & Present ($r = 1.3837$) \\
Static Spheres Pair & Present ($\mathcal{W}=0$) & Present ($\mathcal{W}=0$) & Present ($r_1 = 0.7850, r_2 = 1.2510$) \\
Aschenbach-Like Effect & Present & Present & Present ($\mathrm{d}\Omega/\mathrm{d}r > 0$ on $1.2510<r<1.3837$) \\
\hline
\end{tabular}

\caption{Comparison of static black hole solutions admitting static spheres and Aschenbach-like profiles.}
\label{tab:comparison}
\end{table*}

Regarding structural applications, the conceptual link between static spheres and Dyson-like megastructures is the requirement of static force balance~\cite{13,15,16}. Such megastructures are modelled as swarms of independent statites rather than rigid shells~\cite{16,17}, and applications to compact objects typically involve capturing radiation from accretion disks, coronae, or energetic outflows~\cite{20,21,22,23,24}. Static spheres identify radii where test particles experience zero radial acceleration without orbital motion. However, as noted by Hod~\cite{12_2,13_2}, any real physical structure possesses mass and self-gravity, which backreacts on the metric and shifts the equilibrium radius. Furthermore, the presence of a stable photon sphere is known to host trapped modes that may trigger non-linear instabilities under scalar or tensor perturbations~\cite{14_2}. Assessing the long-term physical stability of these configurations requires specifying a fundamental Lagrangian for the effective source $T^{\rm eff}_{\mu\nu}$.

\section{Conclusions}
In this work, we have analyzed the geodesic structure, static sphere configurations, and Aschenbach-like orbital velocity profiles in a hairy Schwarzschild black hole constructed via Extended Gravitational Decoupling (EGD).

Our main findings are summarized as follows:
\begin{enumerate}
\item The effective energy-momentum tensor $T^{\rm eff}_{\mu\nu}=\alpha\theta_{\mu\nu}$ associated with the EGD deformation violates the Strong Energy Condition (SEC) for $r < 2M - \ell_0$, and satisfies it for $r\ge2M-\ell_0$.
\item For metrics with $g_{tt}g_{rr}=-1$ the identity $(r^{2}\mathcal A')'=16\pi r^{2}p_{t,\mathrm{eff}}$ requires a net negative $\int r^{2}p_{t,\mathrm{eff}}\,\mathrm{d}r$ between the horizon and any static sphere. The innermost static sphere is therefore a maximum of $\mathcal A$, and the SEC is violated locally there when it is unstable, since $p_{t,\mathrm{eff}}=\mathcal A''/16\pi<0$. In the present solution the unstable static sphere ($r_{\mathrm{sp1}}=0.7850$) lies in the SEC-violating region, whereas at the stable one ($r_{\mathrm{sp2}}=1.2510$) $p_{t,\mathrm{eff}}>0$ and the SEC holds. The innermost photon sphere ($r=0.7110$) is also found within the SEC-violating region. For $g_{tt}g_{rr}=-1$ the examples of Ref.~\cite{2_2} violate at least one energy condition, but this has not been proven to be a necessity, so the coincidence should not be read as a proven causal link.
\item The metric supports a pair of static spheres (unstable at $r=0.7850$, stable at $r=1.2510$) with a vanishing net topological charge ($\mathcal{W}=0$), together with three photon spheres with charges $-1,+1,-1$. For $M=\ell_0=1$ both structures appear when $\alpha$ exceeds $e^{2}$, and the three photon spheres persist up to $\alpha\simeq9.26$.
\item Including static spheres as the zero-angular-momentum limit, timelike circular orbits occupy three disconnected branches:
\[
(r_{\mathrm{ph1}},\,r_{\mathrm{sp1}}],
\qquad
[r_{\mathrm{sp2}},\,r_{\mathrm{ph2}}),
\qquad
(r_{\mathrm{ph3}},\,\infty).
\]
Here $r_{\mathrm{ph1}}\simeq0.7110$, $r_{\mathrm{sp1}}\simeq0.7850$, $r_{\mathrm{sp2}}\simeq1.2510$, $r_{\mathrm{ph2}}\simeq1.3837$, and $r_{\mathrm{ph3}}\simeq2.5940$. Photon-sphere endpoints are excluded because $D=0$ there, whereas static-sphere endpoints have $D>0$ and $\mathcal L_z=0$. No circular geodesic exists between the two static spheres. The only marginally stable circular orbit is at $r\simeq5.9151$. The formal zeros of $V''_{\mathrm{eff}}$ at $r\simeq1.1005$ and $1.5671$ lie outside the TCO domain and are not MSCOs.
\item An Aschenbach-like behaviour, $\mathrm{d}\Omega/\mathrm{d}r>0$, occurs for $\ell_0=M$, $\alpha=7.8$ on the branch $1.2510<r<1.3837$ between the stable static sphere and the stable photon sphere; there $\Omega$ is monotonic on each branch.
\item For $\ell_0\le M$ the static-sphere pair is created at $\alpha_{\rm sp}^{*}=e^{2}M/(2M-\ell_0)$ at the radius $r=2M-\ell_0$ where $p_{t,\mathrm{eff}}=0$. The additional photon spheres exist for $\alpha_{\rm lo}<\alpha<\alpha_{\rm hi}$, the edges being given by the roots of the cubic $x^{3}-(3\mu-1)x^{2}+6\mu=0$, and only for $\ell_0>(32-18\sqrt3)M\simeq0.8231M$. This defines four regimes (Fig.~\ref{fig:regime_map}). The case $\ell_0=M$ lies on the degenerate edge where the thresholds $\alpha_{\rm lo}$, $\alpha^{*}_{\rm sp}$, and $\alpha_{\rm SEC}$ all equal $e^{2}$; at the creation of the static-sphere pair the SEC is marginally satisfied.
\item For $g_{tt}g_{rr}=-1$, $\mathrm{d}\Omega_{\mathrm{CO}}^{2}/\mathrm{d}r=(16\pi r\,p_t-3\mathcal A')/(2r^{2})$, so an increasing segment of $\Omega_{\mathrm{CO}}$ requires $p_t>0$. $\Omega_{\mathrm{CO}}$ increases next to every stable photon sphere and next to every stable static sphere~\cite{7}, and, in the subregion I$_\Omega$ of regime I, with neither present. For $\ell_0\le M$ the lower boundary of the parameter region with an increasing segment is $\alpha_\Omega^{*}\simeq3.3093\,\mu$ for $\ell_0\le\ell_{0c}$ and $\alpha_{\rm lo}$ for $\ell_{0c}<\ell_0\le M$.
\item In regime II the connected branch between the inner unstable and the stable photon sphere has an interior minimum of $\Omega_{\mathrm{CO}}$; in regime I$_\Omega$ the single branch has an interior minimum followed by an interior maximum; in regime IV the branch $[r_{\mathrm{sp2}},\infty)$ has an interior maximum and an even number of marginally stable orbits (zero or two in the examples). For $\ell_0=M$, $\alpha=7.8$ (regime III) there are two circular orbits for $\mathcal L_z<\mathcal L_{\rm MSCO}$, three at $\mathcal L_{\rm MSCO}$, and four above it.
\end{enumerate}
These results show that a neutral, asymptotically flat black hole whose only departure from the Schwarzschild solution is a localised geometric deformation reproduces the static spheres, the stable photon sphere, and the Aschenbach-like behaviour reported for electromagnetic and massive-gravity models, and they identify the conditions on $\mathcal A(r)$ under which each of these occurs. The effective source violates the SEC at radii below the one at which the static-sphere pair is created; for the parameters of Sec.~\ref{sec:hsbh} it also violates the NEC in a shell adjoining the horizon, it violates the DEC far from the black hole (Appendix~B), and its fundamental origin is not specified.

\section*{Appendix A: Effective Source of the Metric}
For the line element $\mathrm{d}s^{2}=-\mathcal A\,\mathrm{d}t^{2}+\mathcal A^{-1}\mathrm{d}r^{2}+r^{2}\mathrm{d}\Omega^{2}$ the non-vanishing mixed components of the Einstein tensor are
\begin{align}
G^{t}{}_{t}=G^{r}{}_{r}&=\frac{r\mathcal A'+\mathcal A-1}{r^{2}},\\
G^{\theta}{}_{\theta}=G^{\phi}{}_{\phi}&=\frac{r\mathcal A''+2\mathcal A'}{2r}.
\end{align}
In the EGD construction~\cite{59,60,61} the energy-momentum tensor is split as $\tilde{T}_{\mu\nu} = T^{\rm seed}_{\mu\nu} + \alpha \theta_{\mu\nu}$. For the hairy Schwarzschild solution~\cite{57} the seed is the vacuum Schwarzschild solution, $T^{\rm seed}_{\mu\nu}=0$, so that the whole source is $T^{\rm eff}_{\mu\nu}=\alpha\theta_{\mu\nu}$, and $\mathcal A=1-2M/r+h(r)$ with $h=\alpha e^{-r/\mathcal M}$. The Schwarzschild part therefore drops out of the Einstein tensor:
\begin{align}
G^{t}{}_{t}=G^{r}{}_{r}&=\frac{h+rh'}{r^{2}},\\
G^{\theta}{}_{\theta}&=\frac{rh''+2h'}{2r}.
\end{align}
The effective density and pressures used in the main text are defined by $8\pi T^{{\rm eff}\,\mu}{}_{\nu}\equiv G^{\mu}{}_{\nu}$; the coupling $\alpha$ is contained in $T^{\rm eff}=\alpha\theta$. Eqs.~\eqref{eq:pt_correct} and~\eqref{eq:rho_appB}--\eqref{eq:pt_appB} follow from these components.

\section*{Appendix B: Energy Conditions}
For the metric function $\mathcal{A}(r) = 1 - \frac{2M}{r} + \alpha e^{-r/\mathcal{M}}$ with $\mathcal{M} = M - \ell_0/2$, $\alpha>0$, and $\mathcal M>0$, the effective stress-energy components derived from $G^\mu{}_\nu = 8\pi T^\mu{}_\nu$ (signature $(-+++)$) are given below; Eqs.~(19)--(20) of Ref.~\cite{2_2} give $\rho$ and $p_t$ as $-G^{t}{}_{t}$ and $G^{\theta}{}_{\theta}$ themselves, i.e.\ $8\pi$ times these expressions:
\begin{align}
\rho_{\mathrm{eff}} &= \frac{\alpha e^{-r/\mathcal{M}}}{8\pi \mathcal{M} r^2} (r - \mathcal{M}), \label{eq:rho_appB}\\
p_{r,\mathrm{eff}} &= -\rho_{\mathrm{eff}}, \\
p_{t,\mathrm{eff}} &= \frac{\alpha e^{-r/\mathcal{M}}}{16\pi \mathcal{M}^2 r} \left(r - 2M + \ell_0\right).
\label{eq:pt_appB}
\end{align}
The energy conditions of this source are as follows.

\emph{SEC.} Since $\rho_{\mathrm{eff}}+p_{r,\mathrm{eff}}=0$, the SEC consists of $p_{t,\mathrm{eff}}\ge0$ and $\rho_{\mathrm{eff}}+p_{t,\mathrm{eff}}\ge0$ (the latter is the tangential NEC below). The first requires
\begin{equation}
r \ge 2M - \ell_0 ,
\label{eq:sec_bound_appB}
\end{equation}
and the second is implied by it. For $r<2M-\ell_0$ the SEC is violated; for $M=1,\ell_0=1,\alpha=7.8$ this means $r<1.0$. Outside the horizon, the SEC is violated if and only if $r_h<2M-\ell_0$.

\emph{NEC.} The radial NEC ($\rho_{\mathrm{eff}}+p_{r,\mathrm{eff}}\ge0$) holds trivially, and the tangential NEC reduces to $\rho_{\mathrm{eff}}+p_{t,\mathrm{eff}}=\alpha e^{-r/\mathcal M}(r^{2}-2\mathcal M^{2})/(16\pi\mathcal M^{2}r^{2})\ge0$, i.e.\ to $r^2\ge2\mathcal M^{2}$,
\begin{equation}
r \ge r_{\mathrm{NEC}} \equiv \sqrt{2}\,\mathcal{M} = \sqrt{2}\left(M-\frac{\ell_0}{2}\right).
\end{equation}
For $M=1,\ell_0=1,\alpha=7.8$, $r_{\mathrm{NEC}}\approx0.7071$, which lies between the horizon ($r_h=0.5830$) and the innermost photon sphere ($r=0.7110$). Both radii singled out in the main text, $r_{\mathrm{sp1}}=0.7850$ and the innermost photon sphere at $r=0.7110$, lie outside the shell $r_h<r<r_{\mathrm{NEC}}$ in which the NEC is violated.

\emph{WEC and DEC.} The WEC requires in addition $\rho_{\mathrm{eff}}\ge0$, i.e.\ $r\ge\mathcal M=0.5$, which holds outside the horizon; hence for these parameters the WEC is violated in the same shell as the NEC. The DEC requires $|p_{t,\mathrm{eff}}|\le\rho_{\mathrm{eff}}$ and $|p_{r,\mathrm{eff}}|\le\rho_{\mathrm{eff}}$. It holds for $\sqrt2\,\mathcal M\le r\le(2+\sqrt2)\,\mathcal M$, i.e.\ $0.7071\le r\le1.7071$, and is violated in the shell $r_h<r<r_{\mathrm{NEC}}$ and again far from the black hole, for $r>(2+\sqrt{2})\mathcal{M}\approx1.707$, where $\rho_{\mathrm{eff}}<p_{t,\mathrm{eff}}$. The outer unstable photon sphere at $r=2.5940$ therefore sits in a region where the DEC, though not the SEC or NEC, fails.

\section*{Appendix C: Explicit Forms of the Parametric Functions $\varepsilon_i$}
For completeness, we provide the explicit forms of the parametric functions $\varepsilon_1$, $\varepsilon_2$, $\varepsilon_3$, $\varepsilon_4$, and $\varepsilon_5$, and of the auxiliary function $\varepsilon_0$, in terms of which the second derivative of the effective potential on a TCO, Eq.~\eqref{eq:Vpp}, reads $V''_{\text{eff}} = (-\varepsilon_1 + \varepsilon_2 + \varepsilon_3) / \varepsilon_5$ (this expression coincides identically with Eq.~\eqref{eq:Vpp}):
\begin{align}
\varepsilon_0 &= \frac{r^{3}}{2}+\left(-\frac{5 M}{2}+\frac{3 l_{0}}{4}\right) r^{2} \nonumber\\
&\quad + 7 M \left(M -\frac{l_{0}}{2}\right) r + M \left(M -\frac{l_{0}}{2}\right)^{2}, \\
\varepsilon_1 &= 8 \varepsilon_{0} \alpha r {\mathrm e}^{-\frac{2 r}{2 M -l_{0}}}, \\
\varepsilon_2 &= 12 \left(M -\frac{l_{0}}{2}+\frac{r}{3}\right) \alpha^{2} r^{3} {\mathrm e}^{-\frac{4 r}{2 M -l_{0}}}, \\
\varepsilon_3 &= 48 M \left(M -\frac{l_{0}}{2}\right)^{2} \left(M -\frac{r}{6}\right), \\
\varepsilon_4 &= -\frac{r \alpha \left(M -\frac{l_{0}}{2}+\frac{r}{2}\right) {\mathrm e}^{-\frac{2 r}{2 M -l_{0}}}}{3} \nonumber\\
&\quad + \left(M -\frac{r}{3}\right) \left(M -\frac{l_{0}}{2}\right), \\
\varepsilon_5 &= 12 \left(M -\frac{l_{0}}{2}\right) \varepsilon_{4} r^{3}.
\end{align}

\begin{center}
\textbf{Declaration on the Use of Artificial Intelligence Technologies}
\end{center}
In the preparation of this manuscript, the authors utilized artificial intelligence-based technologies solely for optimizing and refining the linguistic structure, ensuring compliance with academic journal publication standards, and improving language quality, the identification of potential weaknesses,  grammar, and \LaTeX{} typesetting. 
\textbf{All ideas  are entirely original and the direct outcome of the authors' research.} 
It is emphasized that the text refined by these tools was critically and meticulously reviewed multiple times to ensure its alignment with the authors' original intent and scientific findings.\\ \textit{All scientific content, calculations, interpretations, and final decisions have been reviewed, verified, and approved by the authors, who assume full responsibility for this work.}\\\\

\begin{center}
\textbf{The authors declare no conflicts of interest}
\end{center}

\begin{center}
\textbf{Data Availability Statement}
\end{center}
No new data were created or analyzed in this study. Date sharing is not
applicable to this article.
\end{document}